\documentclass[twocolumn]{aastex63}
\usepackage{geometry}
\usepackage{amsmath}
\usepackage{graphicx}
\usepackage{color}

\usepackage{tikz}
\usetikzlibrary{matrix}
\usepackage{natbib}

\usepackage{lineno}

\newcommand{\swift}{\textit{Swift}}

\begin{document}


\title{A non-equipartition shockwave traveling in a dense circumstellar environment around SN\,2020oi}

\author[0000-0002-5936-1156]{Assaf Horesh}
\affiliation{Racah Institute of Physics, The Hebrew University of Jerusalem, Jerusalem 91904, Israel}
\author[0000-0003-0466-3779]{Itai Sfaradi}
\affiliation{Racah Institute of Physics, The Hebrew University of Jerusalem, Jerusalem 91904, Israel}
\author{Mattias Ergon}
\affiliation{Department of Astronomy, The Oskar Klein Center, Stockholm University, AlbaNova, SE-10691 Stockholm, Sweden}
\author{Cristina Barbarino}
\affiliation{Department of Astronomy, The Oskar Klein Center, Stockholm University, AlbaNova, SE-10691 Stockholm, Sweden}
\author[0000-0003-1546-6615]{Jesper Sollerman}
\affiliation{Department of Astronomy, The Oskar Klein Center, Stockholm University, AlbaNova, SE-10691 Stockholm, Sweden}
\author[0000-0002-8079-7608]{Javier Moldon}
\affiliation{Instituto de Astrof\'isica de Andaluc\'ia (IAA, CSIC), Glorieta de las Astronom\'ia, s/n, E-18008 Granada, Spain}
\affiliation{Jodrell Bank Centre for Astrophysics, School of Physics and Astronomy, University of Manchester, Manchester M13 9PL, UK}
\author{Dougal Dobie}
\affiliation{Sydney Institute for Astronomy, School of Physics, University of Sydney, NSW 2006, Australia}
\affiliation{ATNF, CSIRO Astronomy and Space Science, PO Box 76, Epping, NSW 1710, Australia}
\author{Steve Schulze}
\affiliation{Department of Particle Physics and Astrophysics, Weizmann Institute of Science, Israel}
\author{Miguel P\'erez-Torres}
\affiliation{Instituto de Astrof\'isica de Andaluc\'ia (IAA, CSIC), Glorieta de las Astronom\'ia, s/n, E-18008 Granada, Spain}
\author{David R. A. Williams}
\affiliation{Jodrell Bank Centre for Astrophysics, School of Physics and Astronomy, The University of Manchester, Manchester, M13 9PL, UK}
\affiliation{Astrophysics, Department of Physics, University of Oxford, Keble Road, Oxford, OX1 3RH, UK}
\author{Christoffer Fremling}
\affiliation{Caltech Optical Observatories, California Institute of Technology, Pasadena, CA  91125}
\author{Avishay Gal-Yam}
\affiliation{Department of Particle Physics and Astrophysics, Weizmann Institute of Science, Israel}
\author{Shrinivas R. Kulkarni}
\affiliation{Cahill Center for Astrophysics, California Institute of Technology, 1200 E. California Blvd. Pasadena, CA 91125, USA}
\author{Andrew O'Brien}
\affiliation{Sydney Institute for Astronomy, School of Physics, University of Sydney, NSW 2006, Australia}
\author{Peter Lundqvist}
\affiliation{Department of Astronomy, The Oskar Klein Center, Stockholm University, AlbaNova, SE-10691 Stockholm, Sweden}
\author{Tara Murphy}
\affiliation{Sydney Institute for Astronomy, School of Physics, University of Sydney, NSW 2006, Australia}
\author{Rob Fender}
\affiliation{Astrophysics, Department of Physics, University of Oxford, Keble Road, Oxford, OX1 3RH, UK}
\author{Justin Belicki}
\affiliation{Caltech Optical Observatories, California Institute of Technology, Pasadena, CA  91125}
\author[0000-0001-8018-5348]{Eric C. Bellm}
\affiliation{DIRAC Institute, Department of Astronomy, University of Washington, 3910 15th Avenue NE, Seattle, WA 98195, USA} 
\author[0000-0002-8262-2924]{Michael W. Coughlin}
\affiliation{School of Physics and Astronomy, University of Minnesota, Minneapolis, Minnesota 55455, USA}
\author{Eran O. Ofek}
\affiliation{Department of Particle Physics and Astrophysics, Weizmann Institute of Science, Israel}
\author[0000-0001-8205-2506]{V. Zach Golkhou}
\affiliation{DIRAC Institute, Department of Astronomy, University of Washington, 3910 15th Avenue NE, Seattle, WA 98195, USA} 
\affiliation{The eScience Institute, University of Washington, Seattle, WA 98195, USA}
\author[0000-0002-3168-0139]{Matthew J. Graham}
\affiliation{Division of Physics, Mathematics, and Astronomy, California Institute of Technology, Pasadena, CA 91125, USA}
\author{Dave A. Green}
\affiliation{Astrophysics Group, Cavendish Laboratory, 19 J. J. Thomson Ave., Cambridge CB3 0HE, UK}
\author[0000-0002-6540-1484]{Thomas Kupfer}
\affiliation{Kavli Institute for Theoretical Physics, University of California, Santa Barbara, CA 93106, USA}
\author[0000-0003-2451-5482]{Russ R. Laher}
\affiliation{IPAC, California Institute of Technology, 1200 E. California Blvd, Pasadena, CA 91125, USA}
\author[0000-0002-8532-9395]{Frank J. Masci}
\affiliation{IPAC, California Institute of Technology, 1200 E. California Blvd, Pasadena, CA 91125, USA}
\author[0000-0001-9515-478X]{A.~A.~Miller}
\affiliation{Center for Interdisciplinary Exploration and Research in Astrophysics (CIERA) and Department of Physics and Astronomy, Northwestern University, 1800 Sherman Road, Evanston, IL 60201, USA}
\affiliation{The Adler Planetarium, Chicago, IL 60605, USA}    
\author[0000-0002-0466-1119]{James D. Neill}
\affiliation{Caltech Optical Observatories, California Institute of Technology, Pasadena, CA  91125}
\author{Yvette Perrott}
\affiliation{School of Chemical and Physical Sciences, Victoria University of Wellington, PO Box 600, Wellington 6140, New Zealand}
\author{Michael Porter}
\affiliation{Caltech Optical Observatories, California Institute of Technology, Pasadena, CA  91125}
\author{Daniel J. Reiley}
\affiliation{Caltech Optical Observatories, California Institute of Technology, Pasadena, CA  91125}
\author{Mickael Rigault}
\affiliation{Université Clermont Auvergne, CNRS/IN2P3, Laboratoire de Physique de Clermont, 63000, Clermont-Ferrand, France}
\author{Hector Rodriguez}
\affiliation{Caltech Optical Observatories, California Institute of Technology, Pasadena, CA  91125}
\author[0000-0001-7648-4142]{Ben Rusholme}
\affiliation{IPAC, California Institute of Technology, 1200 E. California Blvd, Pasadena, CA 91125, USA}
\author[0000-0003-4401-0430]{David L. Shupe}
\affiliation{IPAC, California Institute of Technology, 1200 E. California Blvd, Pasadena, CA 91125, USA}
\author{David Titterington} 
\affiliation{Astrophysics Group, Cavendish Laboratory, 19 J. J. Thomson Ave., Cambridge CB3 0HE, UK}

\begin{abstract}

We report the discovery and panchromatic followup observations of the young Type Ic supernova, SN\,2020oi, in M100, a grand design spiral galaxy at a mere distance of $14$\,Mpc. We followed up with observations at radio, X-ray and optical wavelengths from only a few days to several months after explosion. The optical behaviour of the supernova is similar to those of other normal Type Ic supernovae. The event was not detected in the X-ray band but our radio observation revealed a bright mJy source ($L_{\nu} \approx 1.2 \times 10^{27} {\rm erg\,s}^{-1} {\rm Hz}^{-1}$). Given, the relatively small number of stripped envelope SNe for which radio emission is detectable, we used this opportunity to perform a detailed analysis of the comprehensive radio dataset we obtained. The radio emitting electrons initially experience a phase of inverse Compton cooling which leads to steepening of the spectral index of the radio emission. Our analysis of the cooling frequency points to a large deviation from equipartition at the level of $\epsilon_e/\epsilon_B \gtrsim 200$, similar to a few other cases of stripped envelope SNe. Our modeling of the radio data suggests that the shockwave driven by the SN ejecta into the circumstellar matter (CSM) is  
moving at $\sim 3\times 10^{4}\,{\rm km\,s}^{-1}$. Assuming a constant  mass-loss from the stellar progenitor, we find that the mass-loss rate is $\dot{M} \approx 1.4\times 10^{-4}\,{M}_{\odot}\,{\rm yr}^{-1}$, for an assumed wind velocity of $1000\,{\rm km\,s}^{-1}$. The temporal evolution of the radio emission suggests a radial CSM density structure steeper than the standard  $r^{-2}$.

\end{abstract}

\section{Introduction}
\label{sec:intro}

The various paths leading to a core-collapse supernova (SN), the explosive death of a massive star, are still not fully understood. 
However, thanks to the increasing number of transient discoveries over the last decade, the study and characterization of thousands of SNe have become possible.  
In particular, the discovery of young SNe, a day to a few days after explosion, and panchromatic followup allow to probe the properties of the stellar progenitors.

In the optical, early observations led to the discovery of high-excitation narrow emission lines (also known as flash spectroscopy features; e.g. \citealt{gal-yam_2014}; \citealt{yaron_2017}; \citealt{Groh_2014}; \citealt{SN1983k_1985}). These features likely originate from a dense confined shell of circumstellar material (CSM), ejected by the progenitor star several years only prior to explosion. Early observations of such shells can reveal the composition of the outer envelope of the exploding star, before it is mixed with elements produced in the explosion itself.

Radio observations probe the interaction of the SN ejecta with the CSM. The CSM closest to the star has been deposited by mass-loss processes (e.g. stellar winds or eruptive mass ejections). Thus early radio observations of young SNe provide information on the mass-loss history of the progenitor star in its latest evolutionary stage, leading to the explosion. As the mass-loss process is linked to the star just prior to explosion, understanding it empirically is a key element in the overall quest for our understanding stellar death. 

Early radio observations have already provided some surprises. For example, PTF\,12gzk \citep{ben-Ami_2012} exhibited early faint radio emission that peaked below $10$\,GHz several days after explosion and quickly faded beyond detection when the SN was $\sim 10$\,days old \citep{horesh_apj_2013}. This behaviour was also observed in SN\,2007gr \citep{soderberg_2010} and SN\,2002ap \citep{berger_2002}. It may point to a fast ($0.2-0.3\,c$) shockwave traveling in a low density CSM environment. Clearly it can only be captured if observations are undertaken early enough. Such SNe, with relatively high shockwave velocities, may represent an understudied population of SNe that link normal Type Ic SNe to relativistic ones. Early radio observations may also play a role studying new types of transients. For instance, early observations of SN\,2018cow \citep{ho_2019}, an optical fast blue transient, revealed a bright plateau of millimeter-wave (mm) emission. The early behaviour of the mm-emission is still not well understood, especially when compared to the radio emission at cm-wavelength, that may be explained by a decelerating circumstellar shockwave (\citealt{margutti_2019}; Horesh et al., in prep).

While the study of the recent mass-loss history from massive stars, via radio (and sometimes also X-ray) observations, is important for all types of SNe, those of stripped envelope SNe (of spectral Types IIb/Ib/Ic ) are of particular interest. These SNe must have undergone enhanced mass-loss in order to lose most of their hydrogen, and in some cases also helium, envelopes. Radio emission has been detected from a number of nearby stripped envelope SNe (e.g. SN\,2004cc, SN\,2007bg, SN\,1990B, SN\,1994I, and SN\,2003L \citealt{wellons_2012}; \citealt{salas_2013}; \citealt{chevalier_fransson_2006} and references therein).
A comprehensive view of the ongoing processes in the SN-CSM shockwave can be obtained if X-ray and optical data are combined with radio measurements. Early combined radio to X-ray observations of SN\,2011dh (\citealt{soderberg_2012}; \citealt{horesh_mnras_2013}) pointed towards a large deviation from equipartition between the shockwave accelerated electron energy and the shockwave enhanced magnetic field energy. Other examples include SN\,2012aw \citep{yadav_2014} in which the steep radio spectrum observed early on showed a significant inverse Compton cooling at frequencies above $1$\,GHz, and SN\,2013df \citep{kamble_2016} that also showed signs of electron cooling by inverse Compton scattering in the radio band. The inverse Compton scattering process in these SNe resulted in enhanced X-ray emission. In both SNe, large deviations from equipartition was found (by a factor of $\sim 200$).

The past observations show the considerable diagnostic value resulting from radio observations of young SNe.
Here, we report the optical discovery of SN\,2020oi, a nearby stripped envelope SN of Type Ic (\S\,\ref{sec:optical_obs}). We conducted a comprehensive radio observing campaign of SN\,2020oi with various facilities (\S\,\ref{sec:radio_obs}) and also obtained X-ray measurements with the {\it Swift} satellite (\S\,\ref{sec:xray_obs}). We present our detailed analysis of the radio measurements in (\S\,\ref{sec:radio_analysis}) and conclude in \S\,\ref{sec:conclusions}.

\section{Optical Observations}
\label{sec:optical_obs}

\subsection{Initial Discovery and Observations}

Supernova SN\,2020oi (a.k.a.\ ZTF\,20aaelulu) in M100 (NGC\,4321; at a distance of $\approx 14$\,Mpc; see \S\,\ref{sec:optical_reduction}) was discovered in $r$-band images obtained by the Zwicky Transient Facility (ZTF; \citealt{bellm,graham,Dekany_2020}) on 2020 January 7 at coordinates $\alpha=12^{h}22^{m}54.93^{s}$, $\delta=+15\degr49\arcmin25.1\arcsec$ (J2000.0). It was initially reported to the Transient Name Server (TNS\footnote{https://wis-tns.weizmann.ac.il/}) by the Automatic Learning for the Rapid Classification of Events (ALeRCE) transient broker service  \citep{2020TNSTR..67....1F},  
which feeds off 
the ZTF  public data stream \citep{patterson_2019}. 
These authors 
noted that SN\,2020oi exhibited a fast rising light curve with an initial $r$-band magnitude of $17.3\pm0.04$.  Inspection of the ZTF 
partnership 
survey data \citep{bellmb} showed that SN\,2020oi was also detected in the 
$g$-band 
by the Palomar $48$-inch (P48) telescope on Julian Date (JD) 2458855.9588, a few hours before the first reported $r$-band detection. As noted in \cite{2020TNSTR..67....1F}, a non-detection limit of $20.5$ in 
$r$-band
was obtained at the position of SN\,2020oi by the P48 telescope on 2020 January 4, $2.9$\,days prior to first detection. 
Spectroscopic observations 
carried out using the SOAR telescope revealed that SN\,2020oi is a Type Ic supernova \citep{2020TNSCR..90....1S}. Thus, SN\,2020oi is one of the most nearby stripped-envelope supernovae in the past decade.

Upon discovery, we 
triggered 
photometric 
observations with the Las Cumbres Observatory (LCO) telescope network. In addition to photometric observations, we carried out $16$ spectroscopic observations of SN\,2020oi over the first three months after discovery using multiple telescopes: the P60 telescope 
\citep[equipped with SEDM;][]{nadiasedm}, 
the Nordic Optical Telescope (NOT), the Palomar $200$-inch telescope (P200) and the Keck Telescope (KECK). The log of these spectral observations is provided in Table~\ref{tab:opticalspec} (see also TNS reports on additional photometric measurements by 
ATLAS, Pan-STARRS and Gaia).


\subsection{Data Reduction and Analysis}
\label{sec:optical_reduction}

In the following, we adopt the time of explosion (or time of first light) as the mid-point between last non-detection (3\,days prior to detection) and first detection, and use the same time window for the uncertainty on the explosion time. Throughout the paper we thus assume 
that 
SN\,2020oi exploded on JD $2458854.50\pm1.46$ (UT 2020 January 06). 

The host galaxy M100 has many redshift-independent distance measurements cataloged on NED\footnote{https://ned.ipac.caltech.edu}, and in this paper we adopt a distance of 14 Mpc corresponding to a distance modulus of $30.72\pm0.06$ mag. According to \cite{Schlafly11} the Milky Way extinction in the direction of M100 is 
E(B$-$V) = 0.023 mag, which we will adopt here. As described at the end of the section, we estimate the host extinction to 0.13 mag, giving a total E(B$-$V) of 0.153 mag.

\begin{figure}
\begin{center}
\includegraphics[width=\linewidth]{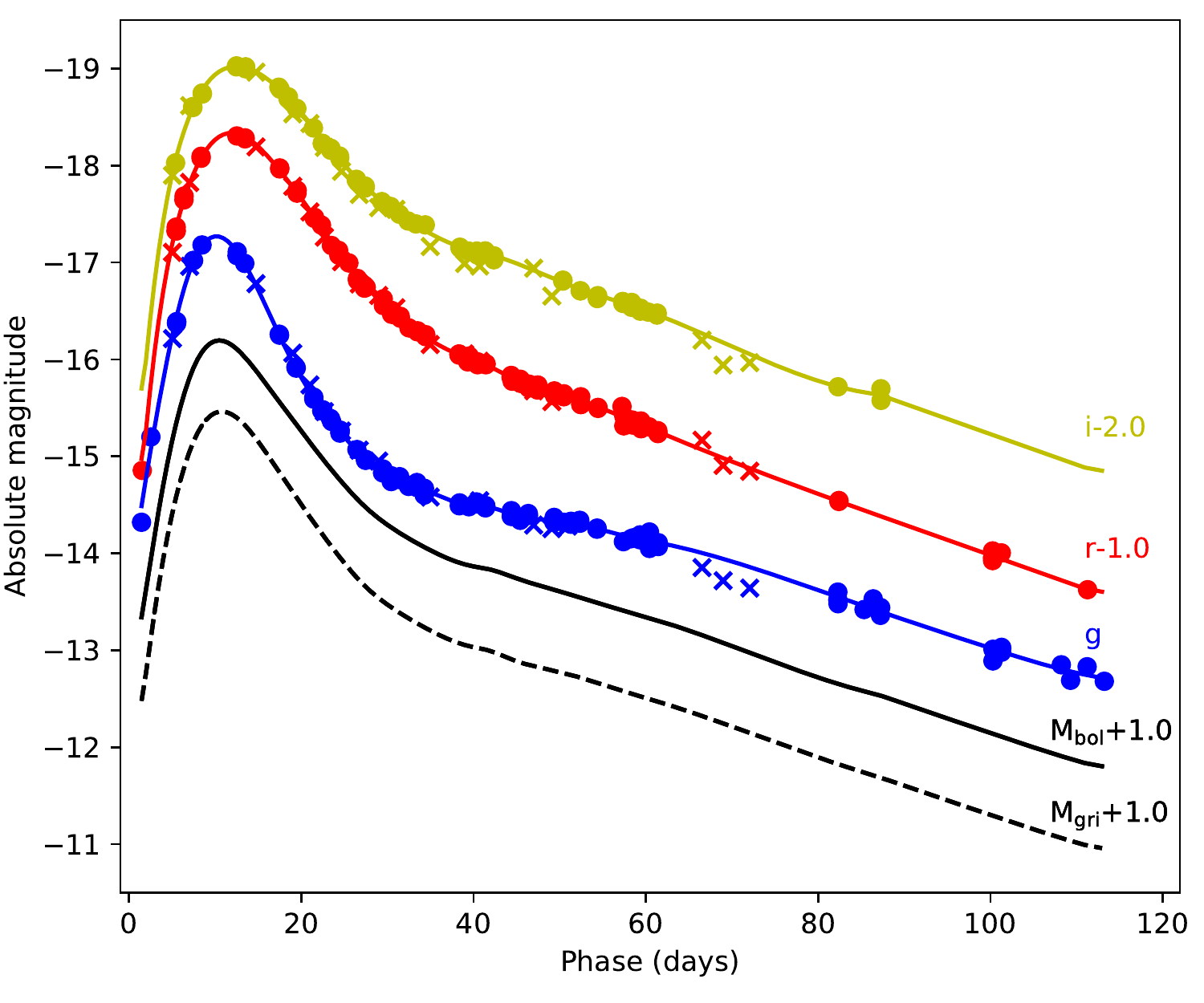}
\caption{\footnotesize{Absolute 
SDSS 
$g$- (blue), $r$- (red) and $i$-band (yellow) light-curves as well as bolometric (black solid line) and $gri$ pseudo-bolometric (black dashed line) light-curves. P48 data is shown as filled circles, LCO data as crosses, and spline fits to the P48 data as solid lines.}}
\label{fig:opticalLC}
\end{center}
\end{figure}

In Fig.\ \ref{fig:opticalLC} we show the absolute 
SDSS
$g$-, $r$- and $i$-band light-curves as well as spline fits to the P48 data. The P48 photometry was reduced with the ZTF production pipeline \citep{Masci2019}, using image subtraction 
based on the \cite{zackay_2016} algorithm. The LCO photometry was reduced with the pipeline described in \cite{2016A&A...593A..68F}, using image-subtraction with template images from the Sloan Digital Sky Survey (SDSS; \citealp{2014ApJS..211...17A}).
In Fig.~\ref{fig:opticalLC} we also show the $gri$ pseudo-bolometric light-curve, calculated from the spline fits to the P48 broad-band light-curves using the method by \citet{Erg14}, as well as the bolometric light-curve
using the bolometric corrections by \citet{lym14}. 

From the spline fits to the P48 broad-band light-curves we measure rise times 
to peak 
$t_{g} = 10.2$ days, $t_{r} = 11.9$ days and $t_{i} = 12.5$ days, peak absolute magnitudes $M_{g} = -17.3$ mag, $M_{r} = -17.3$ mag and $M_{i} = -17.0$ mag, and decline rates from the peak $\Delta M_{g,15} = 2.09$ mag, $\Delta M_{r,15} = 1.53$ mag and $\Delta M_{i,15} = 1.24$ mag.

From the $gri$ pseudo-bolometric and bolometric light-curves we measure rise times 
to peak 
$t_{gri} = 10.8$ days and $t_{bol} = 10.5$ days, peak bolometric magnitudes $M_{gri} = -16.5$ mag and $M_{bol} = - 17.2$ mag, and decline rates from the peak $\Delta M_{gri,15} = 1.64$ mag and $\Delta M_{bol,15} = 1.54$ mag. The $r$-band peak magnitude is within one sigma rms from the average value in the distribution of 44 normal SNe Ic from iPTF (Barbarino et al.\ in prep.). However, SN 2020oi evolves faster than most SNe in this sample, and the $r$-band rise time lies at the lower extreme, and the $r$-band decline rate at the higher extreme of the distribution.

The SEDM Integrated Field Unit (IFU) 
spectra were reduced using pySEDM \citep{Rigault19}, whereas the NOT and P200 spectra were reduced with custom built long slit pipelines \citep{bellmc}. We note that whereas SEDM was primarily constructed to allow classification \citep[see e.g.][]{FremlingRCF}, for this bright nearby supernova the spectral sequence was actually of good quality and also enabled measurements of line velocities.

\begin{figure*}
\begin{center}
\includegraphics[width=0.7\textwidth,angle=90]{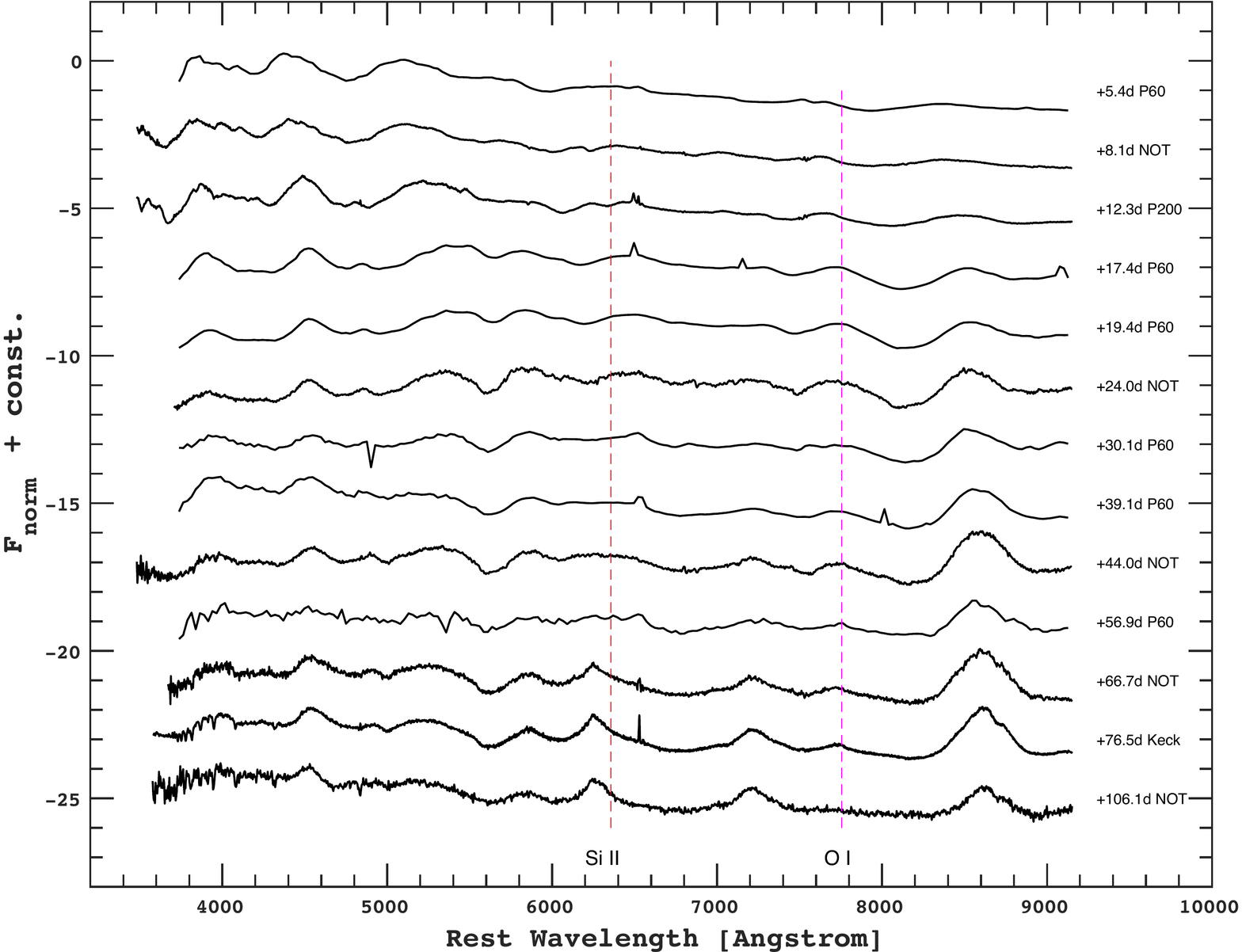}
\caption{\footnotesize{The optical spectral evolution of SN\,2020oi. The phases, in rest-frame days, are provided next to each spectrum.}}
\label{fig:opticalSpec}
\end{center}
\end{figure*}

The sequence of spectra is plotted in Fig.~\ref{fig:opticalSpec}. 
The phases in rest-frame days, with respect to the explosion time, are reported next to each spectrum. At 5 days we measure an absorption minimum velocity and an equivalent with of the \ion{O}{1} 7774 \AA~line of 14\,557~km~s$^{-1}$ and 66 \AA, respectively. At 12 days ($r$-band peak) we measure an absorption minimum velocity and an equivalent width of the \ion{O}{1} 7774 \AA~line of 12\,443 km~s$^{-1}$ and 59 \AA, respectively. Those values are within one sigma rms from the average values in the distribution of 56  normal SNe Ic from PTF and iPTF \citep{Fre18}, although the velocities are on the higher side of the distribution. All observations will be made public via WISeREP\footnote{https://wiserep.weizmann.ac.il},
upon publication.

There appear to be some evidence that the SN exploded in a dense region.
We estimate the reddening in two ways. First, we measure the equivalent width of the Na I D line in the high-quality spectrum from Keck, and obtain $0.74 \pm 0.13$~\AA~where the error is estimated by using multiple choices of the continuum level.  Alternatively, we can attempt to individually measure 0.55 and 0.30~\AA~for Na I D2 and D1 independently (from de-blending two Gaussians).
Using the formalism from \cite{poznanskiEBV}, this provide estimates of E(B-V) = $0.10^{+0.05}_{-0.03}$ mag, and $0.14 \pm 0.05$ mag, respectively.

In addition, we compared the optical colors of SN 2020oi with those of other striped envelope SNe and following the method of 
\cite{stritzingerEBV} this results in E(B-V) = 0.13 mag. Although none of the methods used are precise, 
they are in rough agreement, and in this section we have adopted E(B-V) = 0.13 mag for the host extinction.

\begin{deluxetable}{ccc}
\tablewidth{0pt}
\tabletypesize{\scriptsize}
\tablecaption{Summary of Optical Spectroscopic Observations \label{tab:opticalspec}}
\tablehead{
\colhead{Observation Date} & 
\colhead{Phase} &
\colhead{Telescope+Instrument}  \\
\colhead{(YYYY MMM DD)}  & 
\colhead{(rest-frame days)} &
\colhead{} 
}
\startdata
2020 Jan 11 & 5.4 & P60+SEDM \\
2020 Jan 14 & 8.1 & NOT+ALFOSC \\
2020 Jan 18 & 12.3 & P200+DBSP \\
2020 Jan 23 & 17.4 & P60+SEDM \\
2020 Jan 23 & 17.4 & P60+SEDM \\
2020 Jan 25 & 19.4 & P60+SEDM \\
2020 Jan 29 & 23.3 & P60+SEDM \\
2020 Jan 29 & 24.0 & NOT+ALFOSC \\
2020 Feb 05 & 30.1 & P60+SEDM \\
2020 Feb 14 & 39.1 & P60+SEDM \\
2020 Feb 18 & 44.0 & NOT+ALFOSC \\
2020 Feb 21 & 46.2 & P60+SEDM \\
2020 Mar 03 & 56.9 & P60+SEDM \\
2020 Mar 12 & 66.7 & NOT+ALFOSC \\
2020 Mar 22 & 76.5 & KECK+LRIS \\
2020 Apr 20 & 106.1 & NOT+ALFOSC \\
\enddata
\end{deluxetable}

\section{Radio Observations}
\label{sec:radio_obs}

Radio observations of SN\,2020oi were rapidly initiated using several facilities, including the Karl G. Jansky Very Large Array (VLA), the Arcminute Microkelvin Imager - Large Array (AMI-LA; \citealt{zwart_2008}; \citealt{hickish_2018}), The Australian Telescope Compact Array (ATCA; \citealt{ATCA_mnrs}) and the e-MERLIN Telescope. A possible radio detection in C band was reported on 2020 January 10 \citep{horesh_initial_2020-1} using the VLA under a public observation undertaken by the National Radio Astronomy Observatory (NRAO). A confirmation of the radio detection was made by the AMI-LA telescope \citep{sfaradi_TNS_2020}. Additional detection using the VLA under a public observation undertaken by the NRAO was made on January 11, 2020, in Q band \citep{horesh_TNS_2020}. We then initiated a radio observing campaign of SN\,2020oi under several director discretionary time (DDT) programs on the following facilities: VLA (PI Horesh); AMI-LA (PI Fender \& Horesh); ATCA (PI Dobie); e-MERLIN (PI Perez-Torres \& Horesh). Below we report the radio observations by each facility, the data reduction process and present the results.

\subsection{The Karl G. Jansky Very Large Array}

We observed the field of SN\,2020oi with the VLA on several epochs starting January 10, 2020. The observations (both under a public NRAO program and under our DDT program VLA/19B-350; PI Horesh) were performed in the C- ($5$\,GHz), X- ($10$\,GHz), Ku- ($15$\,GHz), K- ($22$\,GHz), Ka- ($33$\,GHz), and Q- ($44$\,GHz) bands. The VLA was in its most compact (D) configuration during the observations conducted up until January 28, 2020, and in the more extended C configuration from February 10, 2020 onward. 

 We calibrated the data using the automated VLA calibration pipeline available in the Common Astronomy Software Applications (CASA) package \citep{2007ASPC..376..127M}. Additional flagging was conducted manually when needed. Our primary flux density calibrator was 3C286, while J1215+1654 was used as a phase calibrator. Images of the SN\,2020oi field were produced using the CASA task CLEAN in an interactive mode. Each image was produced using $2$\,GHz bandwidth within the VLA bands, resulting in two images for the C- and X-bands, three images for the Ku-band and four images for the K-, Ka- and Q-bands. We also produced images of the full band data for each epoch.

Most observations showed a source at the phase center, which we fitted with the CASA task IMFIT. The image rms was calculated using the CASA task IMSTAT. A summary of the flux density at different observing time and frequency, for the full band images, are reported in Table~\ref{tab:Observations}. We estimate the error of the peak flux density to be a quadratic sum of the image rms, the error produced by CASA task IMFIT and $10$\,\% calibration error. See the online table for more information.

\subsection{The Arcminute Microkelvin Imager - Large Array}

Radio observations of the field of SN\,2020oi were conducted using the AMI-LA telescope. AMI-LA is a radio  interferometer comprised of eight, 12.8-m diameter, antennas producing 28 baselines which extend from 18-m up to 110-m in length and operates with a 5 GHz bandwidth around a central frequency of 15.5 GHz. The first AMI-LA observation of SN\,2020oi occurred on January 11, 2020, about five days after explosion, for four hours. We then continued monitoring SN\,2020oi with high cadence observations.

Initial data reduction, flagging and calibration of the phase and flux, were carried out using $\tt{reduce \_ dc}$, a customized AMI-LA data reduction software package (e.g. \citealt{perrott_2013}).  Phase calibration was conducted using short interleaved observations of J1215+1654, while daily observations of 3C286 were used for absolute flux calibration. Additional flagging was performed using CASA. Images of the field of SN\,2020oi were produced using CASA task CLEAN in an interactive mode. We fitted the source in the phase center of the images with the CASA task IMFIT, and calculated the image rms with the CASA task IMSTAT. We estimate the error of the peak flux density to be a quadratic sum of the image rms, the error produced by CASA task IMFIT and $5$\,\% calibration error. The flux density at each time and frequency are reported in Table~\ref{tab:Observations}.

\subsection{The e-MERLIN Telescope}

We monitored SN\,2020oi with e-MERLIN\footnote{http://www.e-merlin.ac.uk/} at C-band. Observations were conducted within projects DD9007 and CY10006 and consisted of eight runs between January 13 and March 06, 2020, each observation lasting between 5 to 15 hours. The central frequency was $5.1$\,GHz with a bandwidth of $512$\,MHz divided in $512$ frequency channels. 3C286 and OQ208 were used as amplitude and bandpass calibrators, respectively. The phase calibrator, J1215+1654, was correlated at position $\alpha_{\rm J2000.0}=12^{\rm h} 15^{\rm m} 03\fs9791$ and $\delta_{\rm J2000.0}=16\degr 54\arcmin 37\farcs957$ at a separation of $2.1\,\deg$ from the target, and was detected with a flux density of $0.31$\,Jy.

Data reduction was conducted using the e-MERLIN CASA pipeline\footnote{https://github.com/e-merlin/eMERLINCASApipeline} using version v1.1.16 running on CASA version 5.6.2. Before averaging the data, we applied a phase-shift towards the location of an in-beam source located at $1.8\,\deg$ from the target that was used as a reference source to verify the amplitude calibration stability between epochs. A common model for the phase reference calibrator was used to calibrate and image each run. When possible, a phase self-calibration was conducted on the target with one solution per scan combining all the spectral windows. We produced clean images for the target and the in-beam reference source using $\tt{wsclean}$ \citep{offringa-wsclean-2014} with Briggs weighting using a robust parameter of $0.5$ and a cell size of $8$\,mas. The synthesized beam was almost circular with a width of approximately $40$\,mas. The average flux density of the in-beam reference source is $0.36\pm0.02$~mJy. Results of the measurements are shown in Table~\ref{tab:Observations}. We include a $10\%$ uncertainty to the absolute flux density.

\subsection{The Australian Telescope Compact Array}

We conducted two observations of SN\,2020oi using the 6A configuration of ATCA under a Target-of-Opportunity proposal (CX456, PI: Dobie) on January 11 and 18, 2020. Observations were carried out in the $4\,{\rm cm}$ and $15\,{\rm mm}$ bands, with $2\times2\,$GHz bands centered on $5.5/9$\,GHz and $16.7/21.2$\,GHz respectively. We used observations of PKSB$1934-638$ to determine the flux scale of all observations and the bandpass response of the $4\,{\rm cm}$ observations. The $15\,{\rm mm}$ bandpass response was calculated using observations of $1253-055$. Observations of $1222+216$ were used to calibrate the complex gains of all observations.

The visibility data were reduced using standard MIRIAD \citep{1995ASPC...77..433S} routines. In addition, we performed one round of phase-only self-calibration (using a small number of iterations) on the $15\,{\rm mm}$ data. The data were imaged using the MIRIAD clean task using a threshold of $\sim 8$ times the estimated noise background.

We fit a point source at the phase centre using the MIRIAD imfit task, allowing all parameters to freely vary. A summary of the flux density at the different observing frequencies is reported in Table \ref{tab:Observations}.

\subsection{Background radio emission}
\label{sec:background emission}

Archival radio data of the field of SN\,2020oi is available from the Faint Images of the Radio Sky at Twenty-Centimeters (FIRST; \citealt{becker1994vla}) and the NRAO VLA Sky Survey (NVSS; \citealt{condon1998nrao}) archives. They show several nearby radio sources at $1.4$\,GHz, with the closest ones at $4"$ and $8"$ from the reported position of the SN. These radio sources may present a concern for observations conducted with relatively large synthesized beams, as the emission from them can contaminate the SN position.

\begin{figure*}
\begin{center}
    \includegraphics[width=.4\textwidth]{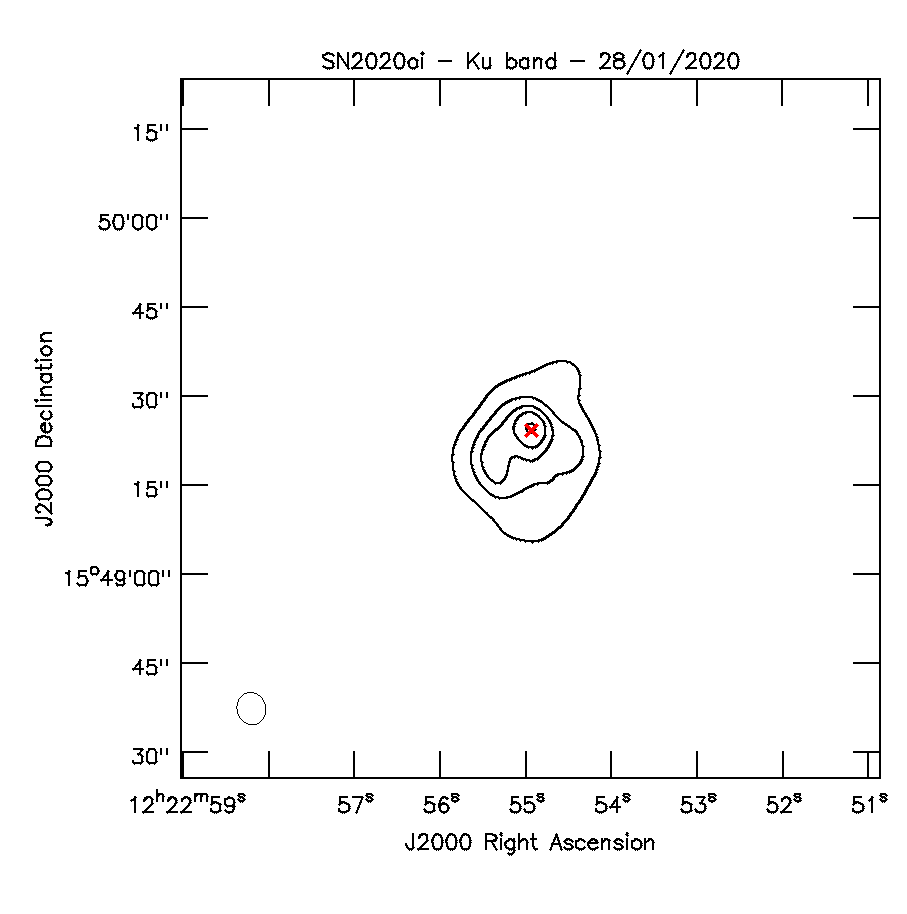}
    \includegraphics[width=.4\textwidth]{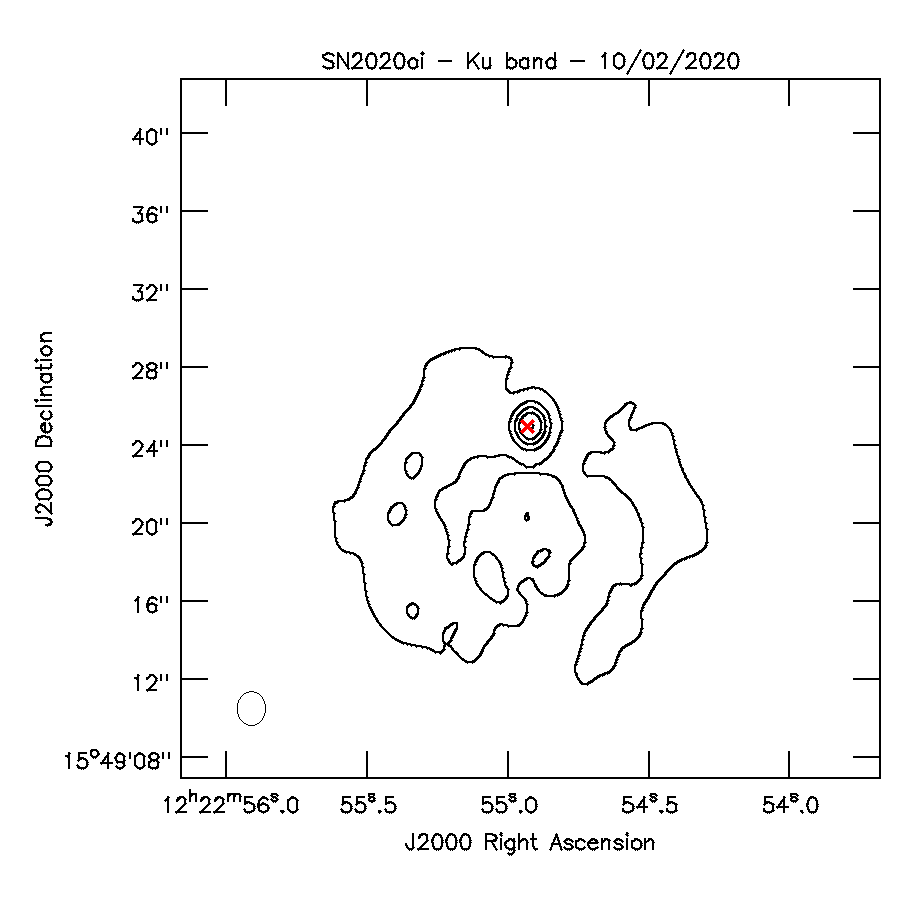}
    \\[\smallskipamount]
    \includegraphics[width=.4\textwidth]{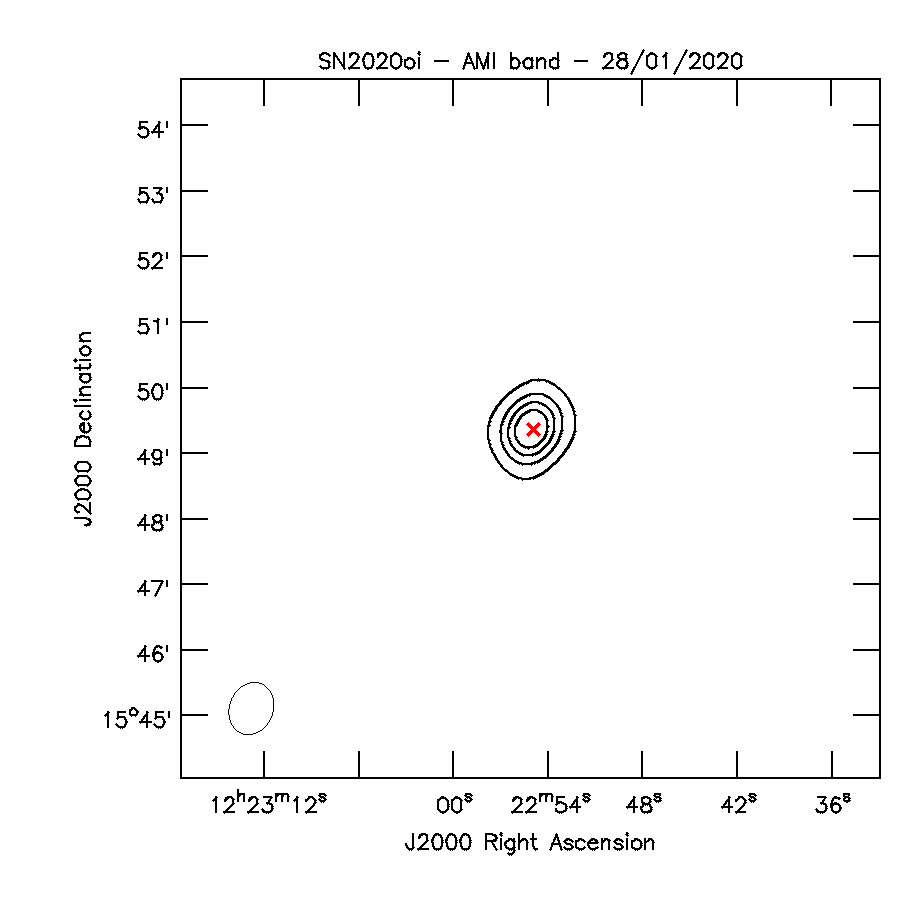}
    \includegraphics[width=.4\textwidth]{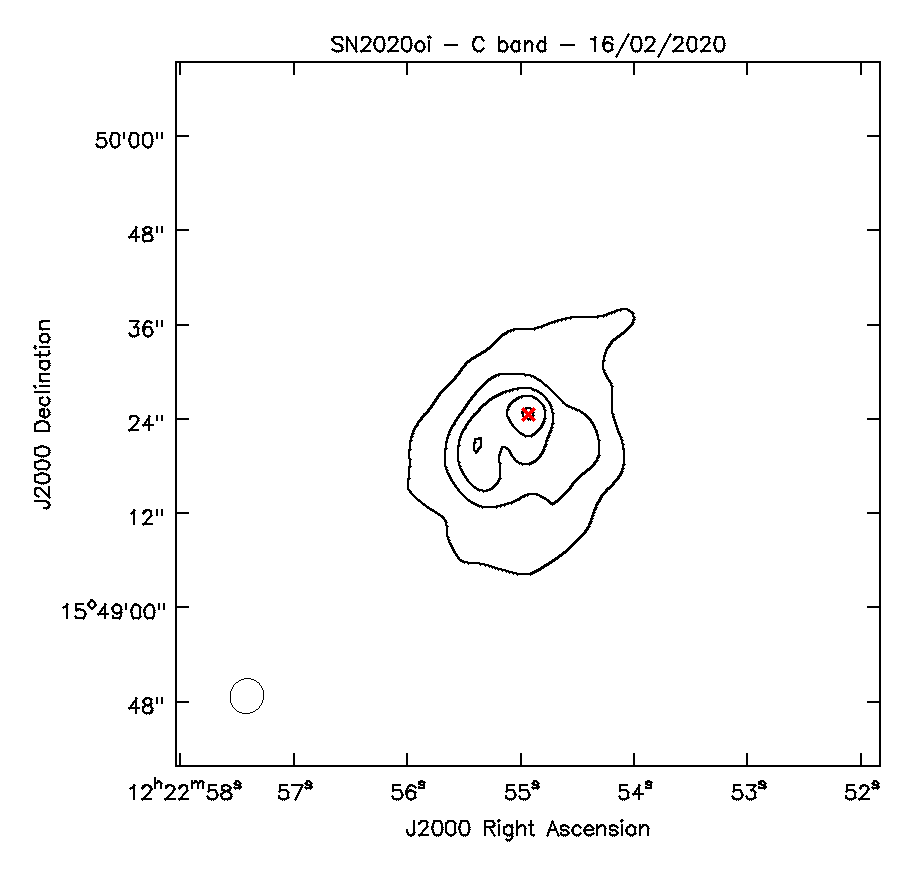}
    \caption{\footnotesize{Contour maps of the field of SN\,2020oi (The SN is marked by a red '$\times$') on different phases and frequencies. These images are examples of how the background contamination discussed in \S\,\ref{sec:background emission} varies according to telescope configuration and the observed band. Top panels show images taken with the VLA in the Ku-band. Left panel is from January 28th, 2020, when the VLA was in D configuration, while the right panel shows an image from February 10th, 2020, when the VLA was in C configuration. The bottom left panel is an AMI-LA $15.5$\,GHz image from January 28th, 2020. The bottom right image is a VLA C-band image from February 16th, 2020. Due to the contamination seen in the Ku-band images taken when the VLA was in  D configuration, and in the VLA C-band images, we do not use these bands (see \S\,\ref{sec:background emission}). The contours represents $10 \%$, $30 \%$, $50 \%$ and $70 \%$ of the peak reported in Table \ref{Table: radio data}.}}
    \label{fig:Images}
\end{center}
\end{figure*}

As described previously, during our observations the VLA changed its configuration from the compact D configuration to the more extended C configuration. Our VLA observations in D configuration had a limited resolution in the Ku-band (lower frequencies were not observed). Hence, we could not resolve the SN emission from the known nearby radio sources. 
This contamination is visible in the upper left panel of Fig. \ref{fig:Images}, showing underlying excess emission in the Ku-band image when the VLA was in D configuration. However, in the upper right panel of this figure, the image of the Ku-band when the VLA was in C configuration shows only negligible contamination. Due to this, flux measurements of Ku-band data taken at D configuration are reported in Table \ref{Table: radio data}, but are not used in our analysis.

Observations conducted in C-band and at the lower sub-band of X-band ($9$\,GHz), when the VLA was in C configuration, are also affected by contamination from the nearby known sources. The lower right image in Fig. \ref{fig:Images} shows the C-band image of SN\,2020oi when the VLA is in C configuration. This image, which exhibits similar features to the ones shown in the Ku-band at D configuration, shows the excessive emission which effects this band. For convenience, we do not show the $9$\,GHz image but this image also shows additional emission at the SN position due to the nearby sources. Due to this, flux measurements below $10$\,GHz that were taken at C configuration are reported in Table \ref{Table: radio data}, but are not used in our analysis.

AMI-LA has a limited resolution at its observing frequency of $15.5$\,GHz due to its short baselines. This results in a large synthetic beam of $\approx 30"$. Hence, we cannot resolve the SN emission from the emission of the known nearby sources and the diffuse emission from the host at any time. As seen in Fig. \ref{fig:Images} we only detect a point source which is comprised of the SN emission and the nearby sources.

Due to the high declination of the source the synthesised beam of ATCA is highly elongated and in the 4\,cm band ($5.5$ and $9$\,GHz) we find it is too elongated to reliably distinguish between emission from the SN and the nearby sources. We therefore do not include these measurements in our analysis.

\LTcapwidth=0.5\textwidth
\begin{longtable}{ccccc}
\caption{SN\,2020oi - radio observations\label{tab:Observations}}
\label{Table: radio data} 
\\ \hline 
$\Delta t$ & Frequency & $F_{\nu}$ & Image RMS & Telescope\\ [1ex]
$[\textrm{Days}]$ & [\textrm{GHz}] & $[\textrm{mJy/beam}]$ & $[\textrm{mJy}]$ & \\ [1ex]
\hline 
\endfirsthead 
\hline 
$\Delta t$ & Frequency & $F_{\nu}$ & Image RMS & Telescope\\ [1ex]
$[\textrm{Days}]$ & [\textrm{GHz}] & $[\textrm{mJy/beam}]$ & $[\textrm{mJy}]$ & \\ [1ex]
\hline 
\endhead 
\hline
\multicolumn{3}{l}{\footnotesize Table \ref{Table: radio data} continues} \\
\hline
\endfoot
\hline
\endlastfoot	
$4.72$ & $15.5$ & $10.42 \pm 0.62$ & $0.07$ & AMI-LA \\  [0.5ex]
$5.15$ & $44$ & $5.12 \pm 0.52$ & $0.03$ & VLA:D \\  [0.5ex]
$5.3$ & $5.5 ^a$ & $0.52 \pm 0.07$ & $0.03$ & ATCA \\  [0.5ex]
$5.3$ & $9 ^a$ & $0.46 \pm 0.05$ & $0.02$ & ATCA \\  [0.5ex]
$5.3$ & $16.7$ & $2.03 \pm 0.21$ & $0.03$ & ATCA \\  [0.5ex]
$5.3$ & $21.2$ & $2.64 \pm 0.27$ & $0.05$ & ATCA \\  [0.5ex]
$6.03$ & $44$ & $4.85 \pm 0.49$ & $0.03$ & VLA:D \\  [0.5ex]
$6.03$ & $33$ & $5.07 \pm 0.51$ & $0.01$ & VLA:D \\  [0.5ex]
$6.03$ & $22$ & $4.22 \pm 0.43$ & $0.01$ & VLA:D \\  [0.5ex]
$6.03$ & $15 ^a$ & $3.04 \pm 0.34$ & $0.01$ & VLA:D \\  [0.5ex]
$6.62$ & $15.5$ & $12.53 \pm 0.68$ & $0.05$ & AMI-LA \\  [0.5ex]
$7.03$ & $44$ & $3.75 \pm 0.38$ & $0.04$ & VLA:D \\  [0.5ex]
$7.03$ & $33$ & $4.37 \pm 0.44$ & $0.01$ & VLA:D \\  [0.5ex]
$7.03$ & $22$ & $4.41 \pm 0.46$ & $0.01$ & VLA:D \\  [0.5ex]
$7.03$ & $15 ^a$ & $3.80 \pm 0.39$ & $0.01$ & VLA:D \\  [0.5ex]
$7.03$ & $5.1$ & $0.22 \pm 0.06$ & $0.04$ & e-MERLIN \\  [0.5ex]
$9.62$ & $15.5$ & $12.8 \pm 0.67$ & $0.06$ & AMI-LA \\  [0.5ex]
$11.02$ & $33$ & $1.84 \pm 0.19$ & $0.02$ & VLA:D \\  [0.5ex]
$11.02$ & $22$ & $2.91 \pm 0.3$ & $0.01$ & VLA:D \\  [0.5ex]
$11.02$ & $15 ^a$ & $4.28 \pm 0.44$ & $0.09$ & VLA:D \\  [0.5ex]
$11.02$ & $5.1$ & $0.67 \pm 0.09$ & $0.04$ & e-MERLIN \\  [0.5ex]
$11.76$ & $15.5$ & $11.85 \pm 0.66$ & $0.05$ & AMI-LA \\  [0.5ex]
$12.72$ & $15.5$ & $11.56 \pm 0.64$ & $0.04$ & AMI-LA \\  [0.5ex]
$12.99$ & $44$ & $0.95 \pm 0.11$ & $0.03$ & VLA:D \\  [0.5ex]
$12.99$ & $33$ & $1.36 \pm 0.15$ & $0.01$ & VLA:D \\  [0.5ex]
$12.99$ & $22$ & $2.21 \pm 0.23$ & $0.01$ & VLA:D \\  [0.5ex]
$12.99$ & $15 ^a$ & $3.40 \pm 0.41$ & $0.09$ & VLA:D \\  [0.5ex]
$12.99$ & $5.1$ & $1.16 \pm 0.17$ & $0.09$ & e-MERLIN \\  [0.5ex]
$13.6$ & $15.5$ & $10.97 \pm 0.56$ & $0.05$ & AMI-LA \\  [0.5ex]
$14.6$ & $15.5$ & $10.57 \pm 0.56$ & $0.07$ & AMI-LA \\  [0.5ex]
$16.7$ & $15.5$ & $11.37 \pm 0.59$ & $0.08$ & AMI-LA \\  [0.5ex]
$17.63$ & $15.5$ & $10.52 \pm 0.53$ & $0.08$ & AMI-LA \\  [0.5ex]
$17.98$ & $44$ & $0.48 \pm 0.06$ & $0.03$ & VLA:D \\  [0.5ex]
$17.98$ & $33$ & $0.70 \pm 0.08$ & $0.01$ & VLA:D \\  [0.5ex]
$17.98$ & $22$ & $1.16 \pm 0.14$ & $0.01$ & VLA:D \\  [0.5ex]
$17.98$ & $15 ^a$ & $2.34 \pm 0.25$ & $0.01$ & VLA:D \\  [0.5ex]
$18.62$ & $15.5$ & $9.89 \pm 0.51$ & $0.05$ & AMI-LA \\  [0.5ex]
$19.59$ & $15.5$ & $9.37 \pm 0.5$ & $0.06$ & AMI-LA \\  [0.5ex]
$20.69$ & $15.5$ & $9.40 \pm 0.49$ & $0.06$ & AMI-LA \\  [0.5ex]
$21.56$ & $15.5$ & $9.49 \pm 0.49$ & $0.09$ & AMI-LA \\  [0.5ex]
$21.98$ & $44$ & $0.35 \pm 0.05$ & $0.03$ & VLA:D \\  [0.5ex]
$21.98$ & $33$ & $0.54 \pm 0.07$ & $0.02$ & VLA:D \\  [0.5ex]
$21.98$ & $22$ & $0.97 \pm 0.10$ & $0.01$ & VLA:D \\  [0.5ex]
$21.98$ & $15 ^a$ & $1.80 \pm 0.19$ & $0.01$ & VLA:D \\  [0.5ex]
$23.61$ & $15.5$ & $9.61 \pm 0.49$ & $0.05$ & AMI-LA \\  [0.5ex]
$24.59$ & $15.5$ & $9.12 \pm 0.49$ & $0.06$ & AMI-LA \\  [0.5ex]
$24.71$ & $5.1$ & $2.18 \pm 0.22$ & $0.03$ & e-MERLIN \\  [0.5ex]
$27.57$ & $15.5$ & $8.89 \pm 0.46$ & $0.05$ & AMI-LA \\  [0.5ex]
$29.56$ & $15.5$ & $8.62 \pm 0.47$ & $0.07$ & AMI-LA \\  [0.5ex]
$30.36$ & $5.1$ & $2.30 \pm 0.23$ & $0.02$ & e-MERLIN \\  [0.5ex]
$34.93$ & $44$ & $0.20 \pm 0.05$ & $0.04$ & VLA:C \\  [0.5ex]
$34.93$ & $33$ & $0.29 \pm 0.04$ & $0.02$ & VLA:C \\  [0.5ex]
$34.93$ & $22$ & $0.48 \pm 0.06$ & $0.02$ & VLA:C \\  [0.5ex]
$34.93$ & $15$ & $0.77 \pm 0.08$ & $0.01$ & VLA:C \\  [0.5ex]
$38.36$ & $5.1$ & $2.19 \pm 0.25$ & $0.09$ & e-MERLIN \\  [0.5ex]
$38.55$ & $15.5$ & $9.70 \pm 0.52$ & $0.06$ & AMI-LA \\  [0.5ex]
$40.95$ & $33$ & $0.29 \pm 0.04$ & $0.02$ & VLA:C \\  [0.5ex]
$40.95$ & $22$ & $0.46 \pm 0.05$ & $0.01$ & VLA:C \\  [0.5ex]
$40.95$ & $15$ & $0.67 \pm 0.08$ & $0.01$ & VLA:C \\  [0.5ex]
$40.95$ & $10 ^a$ & $1.12 \pm 0.13$ & $0.01$ & VLA:C \\  [0.5ex]
$40.95$ & $6 ^a$ & $2.78 \pm 0.30$ & $0.03$ & VLA:C \\  [0.5ex]
$41.93$ & $33$ & $0.32 \pm 0.04$ & $0.02$ & VLA:C \\  [0.5ex]
$41.93$ & $22$ & $0.47 \pm 0.05$ & $0.01$ & VLA:C \\  [0.5ex]
$41.93$ & $15$ & $0.71 \pm 0.07$ & $0.01$ & VLA:C \\  [0.5ex]
$41.93$ & $10 ^a$ & $1.07 \pm 0.13$ & $0.01$ & VLA:C \\  [0.5ex]
$41.93$ & $6 ^a$ & $2.62 \pm 0.28$ & $0.02$ & VLA:C \\  [0.5ex]
$45.52$ & $15.5$ & $9.77 \pm 0.52$ & $0.09$ & AMI-LA \\  [0.5ex]
$52.62$ & $15.5$ & $9.17 \pm 0.49$ & $0.06$ & AMI-LA \\  [0.5ex]
$57.5$ & $15.5$ & $9.05 \pm 0.48$ & $0.05$ & AMI-LA \\  [0.5ex]
$59.29$ & $5.1$ & $1.39 \pm 0.18$ & $0.07$ & e-MERLIN \\  [0.5ex]
$60.29$ & $5.1$ & $1.44 \pm 0.16$ & $0.05$ & e-MERLIN \\  [0.5ex]
$62.48$ & $15.5$ & $9.15 \pm 0.51$ & $0.08$ & AMI-LA \\  [0.5ex]
$68.5$ & $15.5$ & $8.57 \pm 0.51$ & $0.05$ & AMI-LA \\  [0.5ex]
$70.79$ & $33$ & $0.20 \pm 0.03$ & $0.01$ & VLA:C \\  [0.5ex]
$70.79$ & $22$ & $0.25 \pm 0.03$ & $0.01$ & VLA:C \\  [0.5ex]
$70.79$ & $15$ & $0.37 \pm 0.05$ & $0.01$ & VLA:C \\  [0.5ex]
$70.79$ & $10 ^a$ & $0.59 \pm 0.09$ & $0.01$ & VLA:C \\  [0.5ex]
$70.79$ & $6 ^a$ & $1.55 \pm 0.17$ & $0.02$ & VLA:C \\  [0.5ex]
$72.55$ & $15.5$ & $10.01 \pm 0.56$ & $0.11$ & AMI-LA \\  [0.5ex]
$93.76$ & $33$ & $0.09 \pm 0.02$ & $0.01$ & VLA:C \\  [0.5ex]
$93.76$ & $22$ & $0.20 \pm 0.03$ & $0.01$ & VLA:C \\  [0.5ex]
$93.76$ & $15$ & $0.27 \pm 0.04$ & $0.01$ & VLA:C \\  [0.5ex]
$93.76$ & $10 ^a$ & $0.47 \pm 0.07$ & $0.01$ & VLA:C \\  [0.5ex]
$93.76$ & $6 ^a$ & $1.44 \pm 0.15$ & $0.02$ & VLA:C \\  [0.5ex]
\hline
\caption{\footnotesize{A summary of the radio observations conducted with the VLA, AMI-LA, e-MERLIN and ATCA. The first possible detection reported in $\S$\ref{sec:radio_obs} is not reported here due to high contamination from the nearby sources (see $\S$\ref{sec:background emission}). The columns from left to right are as follows: Time since explosion in days; Observed central frequency in GHz, frequencies marked with $^a$ are suspected to be contaminated from nearby sources and are not used in our analysis (see \S\,\ref{sec:background emission}); Peak flux density in mJy/beam; Image RMS in mJy; The telescope with which the observation was made. For the VLA observations, the letter after the colon is for the array configuration of the VLA at the time of observation.}}
\end{longtable}

\section{X-ray Observations}
\label{sec:xray_obs}

\subsection{Observations and data reduction}

\swift\ observed SN\,2020oi with its on board X-ray telescope \citep[XRT;][]{Burrows2005a} in the energy range from $0.3$ to $10$\,keV between 8 January and 28 February 2020. \swift\ also observed the field in 2005--2006 and 2019. We omit the 2019 data because they are affected by SN\,2019ehk. We analysed all data with the online tools of the UK \swift\ team\footnote{\href{https://www.swift.ac.uk/user\_objects/}{https://www.swift.ac.uk/user\_objects/}} that use the methods described in \citet{Evans2007a} and \citet{Evans2009a} and the software package HEAsoft \footnote{\href{https://heasarc.gsfc.nasa.gov/docs/software/heasoft/}{https://heasarc.gsfc.nasa.gov/docs/software/heasoft/}} version 6.26.1 \citep{Blackburn1995a}.

\subsection{Results}

To build the light curve of SN\,2020oi and examine whether any other X-ray source is present at the SN position, we stack the data of each observing segment. We detect emission at the SN position in data from 2005--2006 and in the 2020 data sets. The average count rate in 2005--2006 is $0.013\pm0.001~{\rm ct\,s}^{-1}$ and its rms is $0.005~{\rm ct\,s}^{-1}$ (0.3--10\,keV). The count rate of the data from 2020 is comparable. Spectra of the two epochs show no differences to within the errors, corroborating that the same source dominates (the emission), i.e., the SN is not dominating the X-ray emission.

\begin{table}
\caption{\swift/XRT photometry}\label{tab:xrt}
\begin{tabular}{ccc}
\hline
Time & Phase & Flux \\
$\rm{[MJD]}$ & $\rm{[Days]}$ & $[10^{-13}\,{\rm erg\,cm}^{-2}\,{\rm s}^{-1}]$ \\
\hline
58856.96 & $2.96$ & $1.02 \pm 0.98$ \\
58857.96 & $3.96$ & $1.48 \pm 1.56$ \\
58858.96 & $4.96$ & $<2.20$ \\
58859.96 & $5.96$ & $<2.26$ \\
58860.96 & $6.96$ & $<4.48$ \\
58876.96 & $22.96$ & $1.35 \pm 2.85$\\
58887.96 & $33.96$ & $<5.08$ \\
58898.96 & $44.96$ & $2.72 \pm 2.32$\\
58906.96 & $52.96$ & $<4.69$\\
\hline
\end{tabular}
\end{table}

To recover the SN flux, we numerically subtracted the baseline flux. The error of the baseline-subtracted SN data was computed by adding the standard error of the baseline flux in quadrature to the total error. To convert count-rate to flux, we extracted a spectrum of the 2005--2006 data set. The spectrum is adequately described with an absorbed power-law where the two absorption components represent absorption in the Milky Way and in the host galaxy. The Galactic equivalent neutral-hydrogen column density was fixed to $2.14\times10^{20}~{\rm cm}^{-2}$. The best-fit values of the host absorption is $1.7^{+0.7}_{-0.6}~{\rm cm}^{-2}$ and the photon index\footnote{The photon index is defined as $A(E)\propto E^{-\Gamma}$.} $2.9\pm0.3$ (all uncertainties at 90\% confidence; $\chi^2=209$, degrees of freedom $=162$). To convert the count-rate into flux, we use an unabsorbed energy conversion factor of $4\times10^{-11}~{\rm erg\,cm}^{-2}\,{\rm ct}^{-1}$. Table \ref{tab:xrt} summarises the flux measurements. In that table, we report $3\sigma$ limits for epochs, where the debiased flux level is negative. Furthermore, we applied a 1-day binning. As shown in the table, we do not find any significant X-ray emission from SN\,2020oi, with an approximate upper limit of $\lesssim 5 \times 10^{-13}\,{\rm erg\,cm}^{-2}\,{\rm s}^{-1}$ ($L_{X} \lesssim 1.2 \times 10^{40} {\rm erg\,s}^{-1}$).

\section{Radio Data Modeling and Analysis}
\label{sec:radio_analysis}

In the following section we analyse and model the radio data shown in $\S$\ref{sec:radio_obs} using the SN-CSM interaction model described in \citet{chevalier_1981}. Under the Chevalier model the SN ejecta drive a shockwave into the CSM. As a result, electrons are accelerated at the shockwave front and gyrate in the presence of a magnetic field that is amplified by the shockwave. This gives rise to synchrotron emission which is usually observed in radio waves \citep{chevalier_radio_1982}. The synchrotron emission from the SN can also be absorbed, e.g., by synchrotron self absorption (SSA; \citealt{chevalier_1998}) and/or by free-free absorption (FFA; \citealt{weiler_2002}). Past observations have shown that in most cases for Type Ic SNe, SSA is dominant over FFA (e.g. \citealt{chevalier_fransson_2006}) and we thus use the SSA model described by \cite{chevalier_1998} here.

In the CSM shockwave model, the accelerated relativistic electrons have a power-law energy distribution, $N(E) = N_0 E^{-p}$. Under the SSA only model, the flux from the SN in the optically thick regime ($\nu < \nu_a$, where $\nu_a$ is the frequency at which the optical depth is around unity) is described by
\begin{equation}
    S_\nu \propto \frac{\pi R^2}{D^2} B^{-1/2} \nu^{5/2}.
    \label{eq: optically thick}
\end{equation}

Above $\nu_{a}$, in the optically thin regime, the flux is 
\begin{equation}
    S_\nu \propto \frac{4 \pi f R^3}{3 D^2} N_0 B^{(p+1)/2} \nu^{-(p-1)/2},
    \label{eq: optically thin}
\end{equation}
where $R$ is the radius of the radio emitting shell, $B$ is the magnetic field strength, $D$ is the distance to the SN and $f$ is the emission filling factor.

Measurements of the radio emitting shell radius and of the magnetic field at the shockwave front, can be obtained at any given time using the observed radio spectral peak, $F_{\nu_a}$, and peak frequency $\nu_{a}$, at that time (\citealt{chevalier_fransson_2006}). Assuming $p=3$, the radius is given by
\begin{align}
   \nonumber R  = & 4.0 \times 10^{14} \: f_{eB}^{-1/19} \left( \frac{f}{0.5} \right)^{-1/19} \left( \frac{F_{\nu_a}}{\rm{mJy}} \right)^{9/19} 
  \\
    &\left( \frac{D}{\rm{Mpc}} \right)^{18/19} \left( \frac{\nu_a}{5 \: \rm{GHz}} \right)^{-1} \: \rm{cm},
    \label{eq: Radius_chevalier}
\end{align}
and the magnetic field strength is
\begin{align}
    \nonumber B = & 1.1 \: f_{eB}^{-4/19} \left( \frac{f}{0.5} \right)^{-4/19} \left( \frac{F_{\nu_a}}{\rm{mJy}} \right)^{-2/19}
    \\
    &\left( \frac{D}{\rm{Mpc}} \right)^{-4/19} \left( \frac{\nu_a}{5 \: \rm{GHz}} \right) \: \rm{G}
    \label{eq: Magnetic_chevalier}
\end{align}
where $f_{eB}$ is the ratio between the fraction of shock wave energy deposited in relativistic electrons ($\epsilon_e$) and the fraction of shock wave energy converted to magnetic fields ($\epsilon_B$).

In our analysis below, we will model the data in several ways (similar to e.g. \citealt{horesh_mnras_2013}). First, data from observing epochs where a radio peak is present, will be analyzed individually by fitting single epoch spectra models to those measurements. These individual fits will provide single measurements of the shockwave radius, the magnetic field and the power-law of the electron energy distribution. Second, we model the temporal evolution of single frequencies. Then, we also attempt to perform a combined fit of the full dataset that includes also the time evolution of the radio emission. 


\subsection{Single Epoch Spectral Modeling}
\label{sec:single_epoch}

\begin{figure}
\begin{center}
\includegraphics[width=\linewidth]{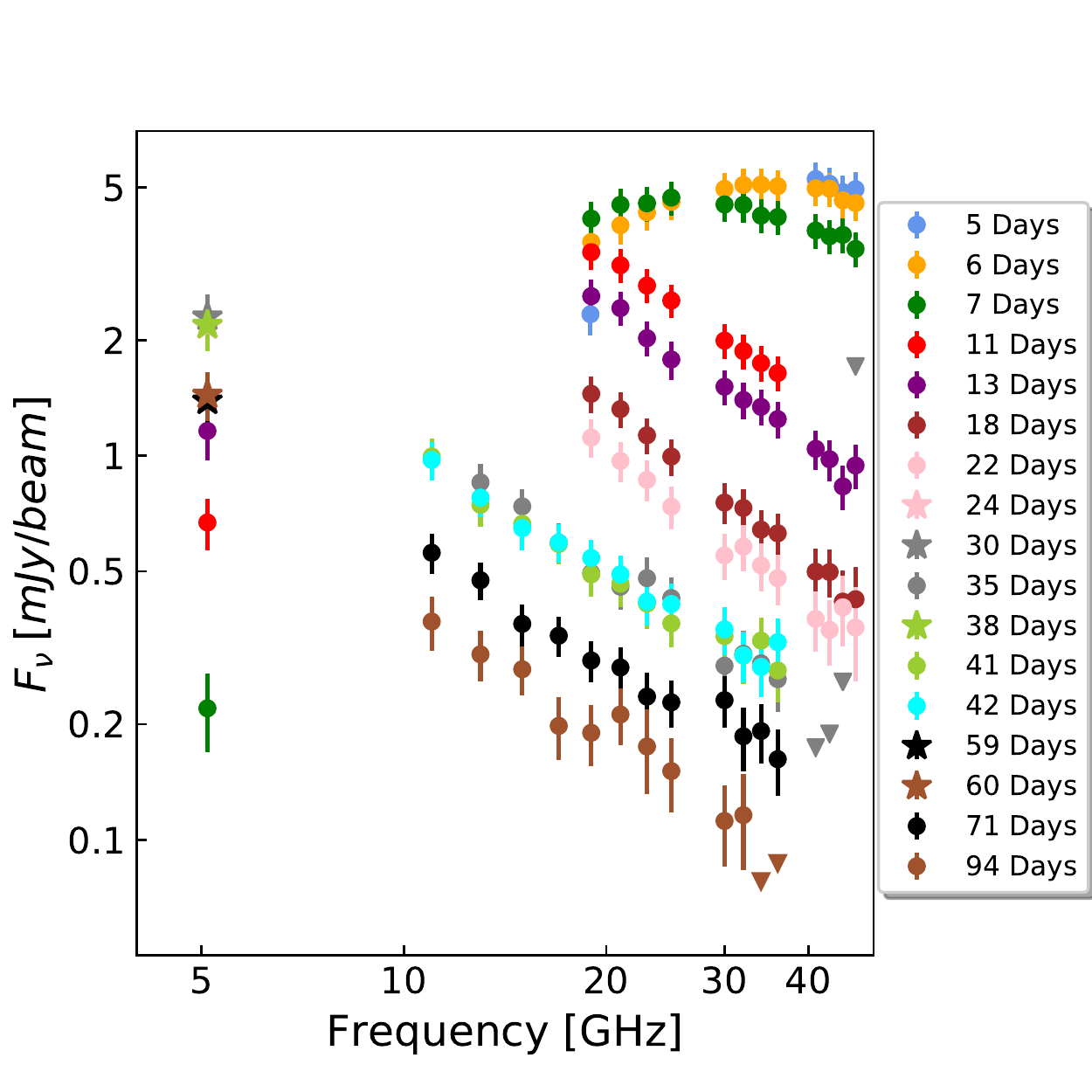}
\caption{\footnotesize{Radio emission as a function of observing frequency, for different epochs. These radio measurements were produced by the VLA, e-MERLIN and ATCA as described in $\S$\ref{sec:radio_obs} and reported in Table \ref{Table: radio data}.}}
\label{fig:VLA Light curves}
\end{center}
\end{figure}

Here we perform individual analysis of the broadband spectral radio data obtained on five to thirteen days after explosion, according to Eq.~\ref{eq: Radius_chevalier} and \ref{eq: Magnetic_chevalier}. The individual spectrum (see online Table) are shown in Fig.~\ref{fig:VLA Light curves}. Note that we do not use the AMI-LA data in our analysis since it suffers from contamination as described in $\S$\ref{sec:background emission} which leads to large uncertainties in the AMI-LA flux measurements. The VLA Ku-band data is also not included in our analysis here (see \S\ref{sec:background emission}). 

\begin{figure}
\begin{center}
\includegraphics[width=\linewidth]{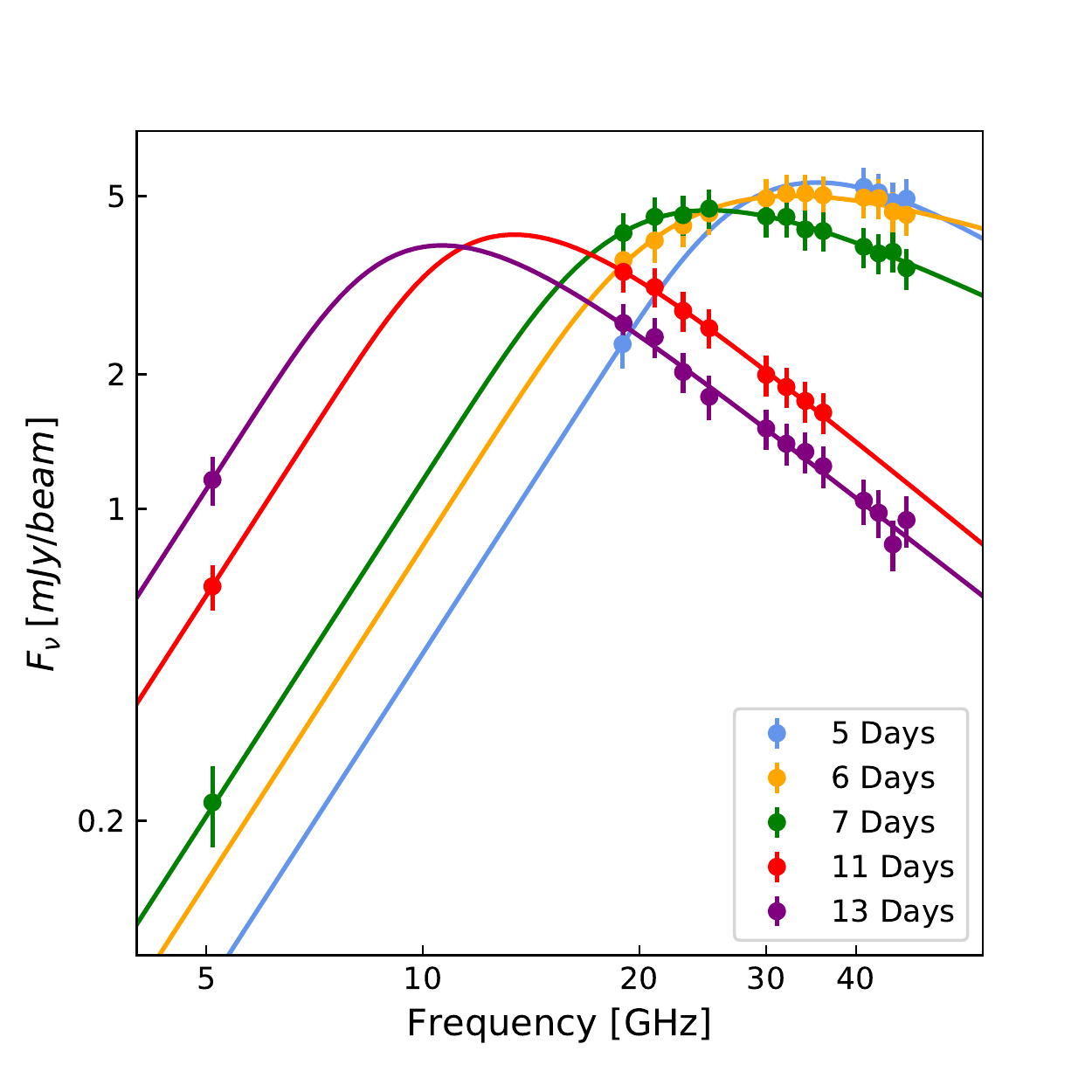}
\caption{\footnotesize{Model fitting of Eq. $4$ in \cite{chevalier_1998} to the observed radio spectra containing information about the spectral peak (using $\chi^2$ minimization fitting to each epoch separately; see \S 5.1).}} 
\label{fig:Chevalier fits - peaks}
\end{center}
\end{figure}

To obtain the spectral peak and frequency, we fit a generalized form of Eq.~$4$ in \cite{chevalier_1998} to the radio spectrum. The free parameters here are $F_{\nu_a} \left( t_a \right)$ and $\nu_a$. 
The spectral index of the optically thin regime, $\beta$, is also a free parameter in the fitted model. Since we fit to a single epoch we use $t=t_a$, where $t$ is the time of observation.

The spectral index of the optically thin synchrotron emission is assumed to be a function of the electron energy power-law index. In the non-cooling regime, the spectral index is defined as $\beta=-(p-1)/2$. However, the estimation of the spectral index $\beta$ and hence $p$ in our case has some limitations. Since in the first three epochs the data do not span a wide enough frequency range after the peak, we probably do not see the radio emission settle fully onto the optically thin regime. Thus, the best fit spectral indexes in these epochs do not represent the real values well enough. In addition, the spectral index of the electron energy density may be effected by electron cooling as discussed in $\S$\ref{sec:spectral_index}. When cooling is present the spectral index will be steeper than what is expected according to the $p$-based relation above. Due to these drawbacks, in our analysis below, we use $p=3$ based on the average value observed in past stripped envelope SNe (e.g., \citealt{chevalier_fransson_2006}; \citealt{soderberg_2012}; \citealt{horesh_mnras_2013}). If $p=3$ then the expected spectral index is $\beta=-1$, which is the value we eventually observe at late times ($\S$\ref{sec:spectral_index}). If  the real $p$ is in the range $2.5<p<3.5$, this will add an additional uncertainty to our analysis below (see Table~\ref{Table: Peak fits}).

The results of the fitting procedure described above are summarized in Table \ref{Table: Peak fits} (including the minimum $\chi^2 _r$, where $\chi^2 _r = \chi^2 /$\,dof, and dof=degrees of freedom), and the observed radio emission together with the best fit models are shown in Fig. \ref{fig:Chevalier fits - peaks}. The radius of the emitting shell and the magnetic field strength are calculated using Eqs. \ref{eq: Radius_chevalier} and \ref{eq: Magnetic_chevalier} assuming equipartition with $\epsilon_e=\epsilon_B=0.1$ (see however the discussion of the effect of non-equipartition on these estimates in \S\ref{sec:non_eq}), and also assuming $p=3$ (Table \ref{Table: Peak fits}). The shockwave velocity, early on, is on average $\approx 4\times 10^{4}\,{\rm km\,s}^{-1}$ (see however \S\,\ref{sec:non_eq}). This velocity represents the velocity of the leading edge of the SN ejecta and thus is expected to be higher by a factor of a few than the photospheric velocities measured using the optical emission which originates from deeper and slower regions within the SN ejecta. The velocity we measure here is a factor of $>2$ higher than the photospheric velocity we measure at early times (\S\,\ref{sec:optical_reduction}).

\begin{figure}
\begin{center}
\includegraphics[width=0.985\linewidth]{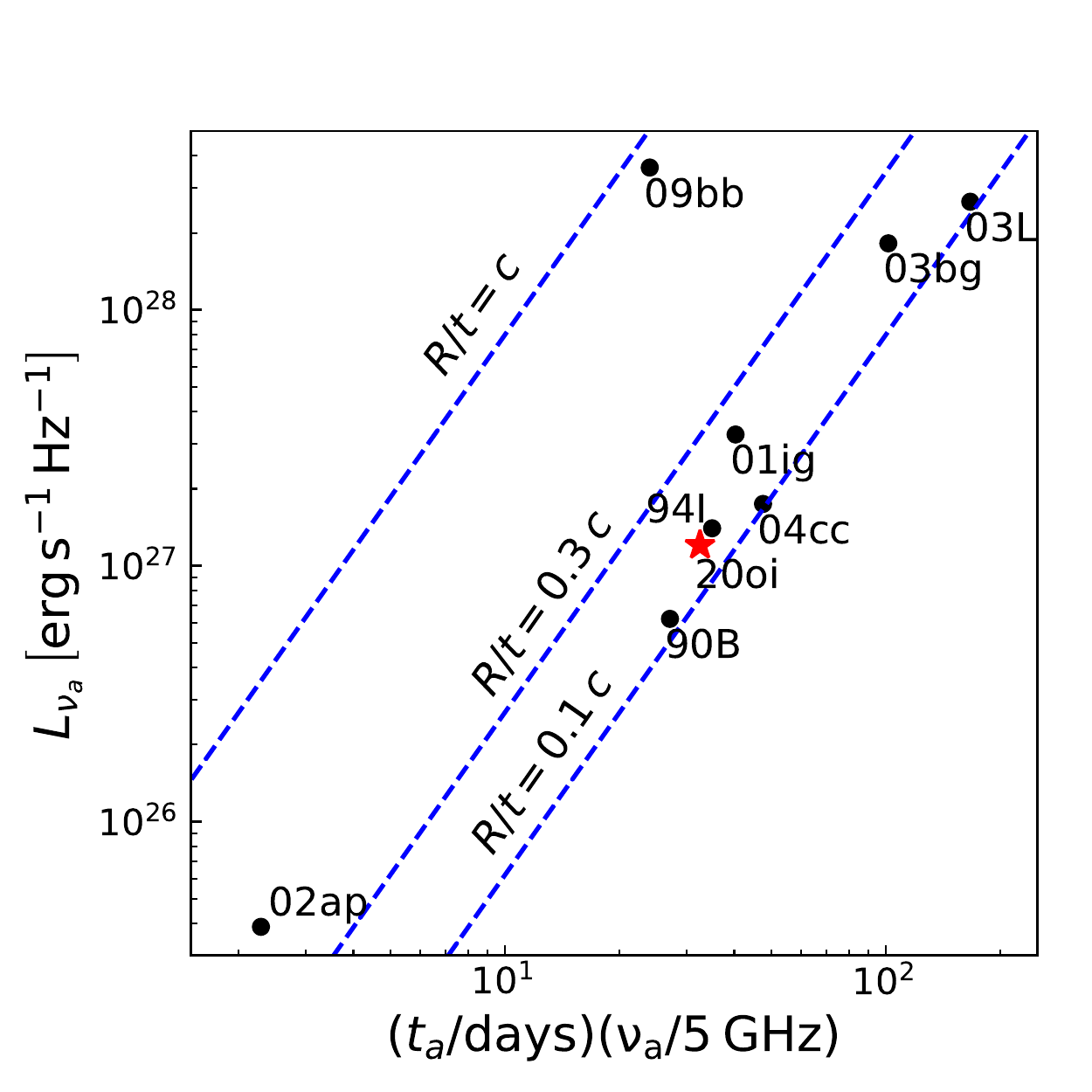}
\caption{\footnotesize{Chevalier's diagram showing the peak luminosity vs. peak frequency and time. Black circles mark other known type Ic SNe. For SN\,1990B, SN\,1994I, SN\,2001ig, SN\,2002ap, SN\,2003L and SN\,2003bg we used the reported values in \cite{chevalier_fransson_2006}. For SN\,2009bb we used the value reported by \cite{soderberg2010relativistic}. For SN\,2004cc we used the value reported by \cite{wellons_2012}. The red star marks SN\,2020oi as fitted to the spectrum, six days after explosion (Table \ref{Table: Peak fits}). Lines of equal $R/t$ are plotted in blue, according to Eq. \ref{eq: Radius_chevalier}.}}
\label{fig:Chevalier-velocities}
\end{center}
\end{figure}

A simple comparison diagnostic tool of the shockwave velocity is the so called Chevalier diagram (e.g.~\citealt{chevalier_1998}). This is a diagram of the measured $L_{\nu_a} - t_a \nu_a$ plane (where $L_{\nu_a}$ is the peak luminosity), that is intersected by diagonal equal velocity lines. We plot the radio peak measurement of SN\,2020oi ($6$\,days after explosion) in Fig. \ref{fig:Chevalier-velocities}, together with a small sample of measurements of other Type Ic SNe. As shown in the figure, the SN-CSM shockwave velocity of SN\,2020oi is quite similar to the ones measured in other normal Type Ic SNe.

We next assume that the CSM originate from stellar winds. We also make the usual assumption that the mass-loss rate and wind velocity were constant on average in the period of time during which the CSM that the SN ejecta interacts with, was created. The CSM in this case has a density structure of $\rho\propto {\dot{M}}/(v_{w} r^{2})$, where ${\dot{M}}$ is the mass-loss rate and $v_{w}$ is the wind velocity. As the energy density of the magnetic energy is a fraction of the shockwave energy density, which depends on the CSM density, we can derive, using the magnetic field strength and the shockwave radius found above, the mass-loss rate
\begin{align}
    \nonumber \dot{M} = & 5.2 \times 10^{-6} \left( \frac{\epsilon_B}{0.1}\right)^{-1} \left( \frac{B}{1 \: \rm{G}} \right)^{2} \\ & \left( \frac{t}{10 \: \rm{days}} \right)^{2} \left( \frac{v_w}{1000 \: {\rm km\,s}^{-1}} \right) \: M_{\odot}\,{\rm yr}^{-1}.
    \label{eq:mass-loss rate}
\end{align}

Estimating the mass-loss rate alone requires an assumption of the wind velocity. Stripped envelope SNe are believed to originate from Wolf-Rayet stars which have fast winds of the order of $v_w = 1000\,{\rm km\,s}^{-1}$  (\citealt{chevalier_fransson_2006}; \citealt{smith_2014}). Adopting the above wind velocity, we estimate the mass-loss rate at various epochs (see Table~\ref{Table: Peak fits}). The average mass-loss rate from the progenitor of SN\,2020oi is $\dot{M}=1.2 \times 10^{-5}$\,$M_{\odot}\,{\rm yr}^{-1}$ (see however \S\,\ref{sec:non_eq}), which is typical to normal Type Ic SNe (e.g. \citealt{chevalier_fransson_2006}).

\LTcapwidth=1.01\textwidth
\begin{longtable*}{ccccccccc}
\caption{SN2020oi - Radio spectral fits and the derived equipartition shockwave parameters} 

 \label{Table: Peak fits}
 \\ \hline
 $t_a$ & $F_{\nu_a} \left( t_a \right)$ & $\nu_a$ & $\beta$ & $\chi^2 _r$ (dof) & $R$ & $B$ & $v_{sh}$ & $\dot{M}$ \\ [1ex]
 $[\textrm{Days}]$ & $[\textrm{mJy/beam}]$ & [\textrm{GHz}] & & & [\textrm{$10^{15}$ cm}] & [\textrm{G}] & [\textrm{$10^4\,{\rm km\,s}^{-1}$}] & \textrm{$\left[ 10^{-5} \: M_{\odot}\,{\rm yr}^{-1} \right]$} \\ [1ex]
 \hline
 \endfirsthead
 \endhead
 \hline
 \endfoot
 
\hline
 \endlastfoot	
$5$ & $5.24 \pm 0.51$ & $31.4 \pm 1.8$ & $-1.04 \pm 0.31$ & $0.10 \, (1)$ & $1.70 \pm 0.26 $ & $3.33 \pm 0.61$ & $3.9 \pm 1.3$ & $1.4 \pm 1.0$ \\  [0.5ex]
$6$ & $4.51 \pm 0.61$ & $23.7 \pm 1.7$ & $-0.53 \pm 0.37$ & $0.06 \, (8)$ & $2.10 \pm 0.35 $ & $2.55 \pm 0.53$ & $4.0 \pm 1.2$ & $1.2 \pm 0.8$ \\  [0.5ex]
$7$ & $4.40 \pm 0.84$ & $20.4 \pm 1.7$ & $-0.77 \pm 0.39$ & $0.04 \, (9)$ & $2.41 \pm 0.44 $ & $2.21 \pm 0.56$ & $4.0 \pm 1.1$ & $1.3 \pm 0.8$ \\  [0.5ex]
$11$ & $4.07 \pm 0.78$ & $12.6 \pm 0.6$ & $-1.31 \pm 0.15$ & $0.02 \, (5)$ & $3.76 \pm 0.64 $ & $1.37 \pm 0.33$ & $4.0 \pm 0.9$ & $1.2 \pm 0.7$ \\  [0.5ex]
$13$ & $3.84 \pm 0.39$ & $9.89 \pm 0.38$ & $-1.25 \pm 0.12$ & $0.19 \, (9)$ & $4.66 \pm 0.83 $ & $1.08 \pm 0.20$ & $4.1 \pm 0.9$ & $1.0 \pm 0.6$ \\  [0.5ex]
$20.4 \pm 1.8$ & $2.21 \pm 0.37$ & $5.1$ & - & $0.69 \, (3)$ & $6.96 \pm 1.70 $ & $0.59 \pm 0.16$ & $3.9 \pm 1.0$ & $0.8 \pm 0.4$ \\  [0.5ex]
\hline
\caption{\footnotesize{The first five rows show a summary of the radio peak fit (using Eq. $4$ in \citealt{chevalier_1998} and $t=t_a$) at five separate epochs and the derived measurements of the shockwave radius , the magnetic field strength (Eq.~\ref{eq: Radius_chevalier} and ~\ref{eq: Magnetic_chevalier} respectively, assuming equipartition and $f=0.5$), the shockwave velocity (assuming $v_{sh}=R/t_a$) and, the mass-loss rate is calculated using Eq. \ref{eq:mass-loss rate} assuming wind velocity of $v_w=1000\,{\rm km\,s}^{-1}$ and $\epsilon_B = 0.1$. The last row shows a summary of the radio peak fit (using Eq. $4$ in \citealt{chevalier_1998} and $\nu=\nu_a$) for the $5.1$\,GHz light curve measured by e-MERLIN. As we note above the radius and magnetic field here are calculated assuming $p=3$. An uncertainty of $\Delta p=\pm 0.5$ will result in additional uncertainty of $17\%$ and $27\%$ in $R$ (and thus also in $v_{sh}$) and $B$, respectively. Note, that a deviation from equipartition (as the one we discuss below) will significantly change the estimates of $R$, $v_{sh}$, and $B$.}}
\end{longtable*}

\subsection{Single Frequency Temporal Analysis}
\label{sec:single_frequ}

Here we perform analysis of the time evolution for individual observed frequencies. At $5.1$\,GHz, the e-MERLIN observations cover both the rising and declining phases in the light curve. Thus, we model the transition from optically thick to optically thin emission at this frequency. Due to the early peak time in the higher observed frequencies, the VLA single frequency light curves does not span a wide enough range for a full light curve fit. Therefore, for those frequencies we only fit a power law in time for the decaying light curves observed by the VLA. Finally, we use the time evolution power laws obtained by the VLA light curves to evaluate the excessive emission contaminating the flux measurements obtained by AMI-LA.

The temporal evolution of the radio emission in both the optically thick and thin regimes is modeled with power-laws. \cite{chevalier_1998} model $1$ assumes $R \sim t^m$ and $B \sim t^{-1}$ and that the flux temporal evolution is $F_{\nu} \sim t^{b}$ where $b = - \left( p + 5 - 6m \right)/2$ (optically thin regime) and $F_{\nu} \sim t^{a}$ where $a = 2m + 0.5 $ (optically thick regime). These definitions of $a$ and $b$ are valid if electron cooling does not effect the emission in the observed frequency.

We fit a generalized form of Eq.\ $4$ in \cite{chevalier_1998} to the $5.1$\,GHz radio light curve measured by e-MERLIN. The free parameters here are $F_{\nu_a} \left( t_a \right)$, $t_a$, and the temporal power-law indexes $a$ and $b$, defined above. Since we fit a single frequency light curve, we use $\nu=\nu_a$, where $\nu$ is the observed frequency. The resulted power law index are $a=2.60 \pm 0.32$ and  $b=-0.81 \pm 0.17$. The fitted peak flux at $5.1$\,GHz is $F_{\nu_a} = 2.21 \pm 0.37$\,mJy, at $20.4 \pm 1.8$\,days after explosion ($\chi^2 _r = 0.69$, dof$=3$). We used the peak at $5.1$\,GHz to calculate $R$, $B$, $v_{sh}$ and $\dot{M}$ (we assume $p=3$, $v_{sh} = R/t_a$ and $v_w=1000\,{\rm km\,s}^{-1}$) and we report them in Table \ref{Table: Peak fits}.

We now examine the time evolution of the radio emission as it is manifested in the VLA K-, Ka- and Q- bands. We use $\chi^2$ minimization to fit a power law to the declining regimes of the light curves, starting $11$\,days after explosion since the flux at earlier epochs is around peak for all observed bands. We divide our analysis into two regimes, first we fit for times $\leq22$\,days since our analysis suggests that electron cooling takes place up to this time (see $\S$\ref{sec:spectral_index}). We then fit a power law for times $\geq35$\,days as electron cooling is not significant anymore. The fitting results in K-band and Ka-band using data from $11$ to $22$\,days after explosion are $b=-1.67 \pm 0.14$ ($\chi^2 _r= 0.97$, dof$=1$) and $b=-1.83 \pm 0.10$ ($\chi^2 _r= 0.44$, dof$=1$), respectively. When using data from $35$ to $94$\,days after explosion the power law in the K-band is $b=-0.97 \pm 0.10$ ($\chi^2 _r= 0.63$, dof$=2$), while in the Ka-band we get $b=-1.03 \pm 0.26$ ($\chi^2 _r= 3.33$, dof$=2$). The fit of Q-band data from $11$ to $22$\,days gives $b=-1.37 \pm 0.12$ ($\chi^2 _r= 2.2$, dof$=2$). This band was observed again only $35$\,days after explosion and was not detected.

\begin{figure}
\begin{center}
\includegraphics[width=\linewidth]{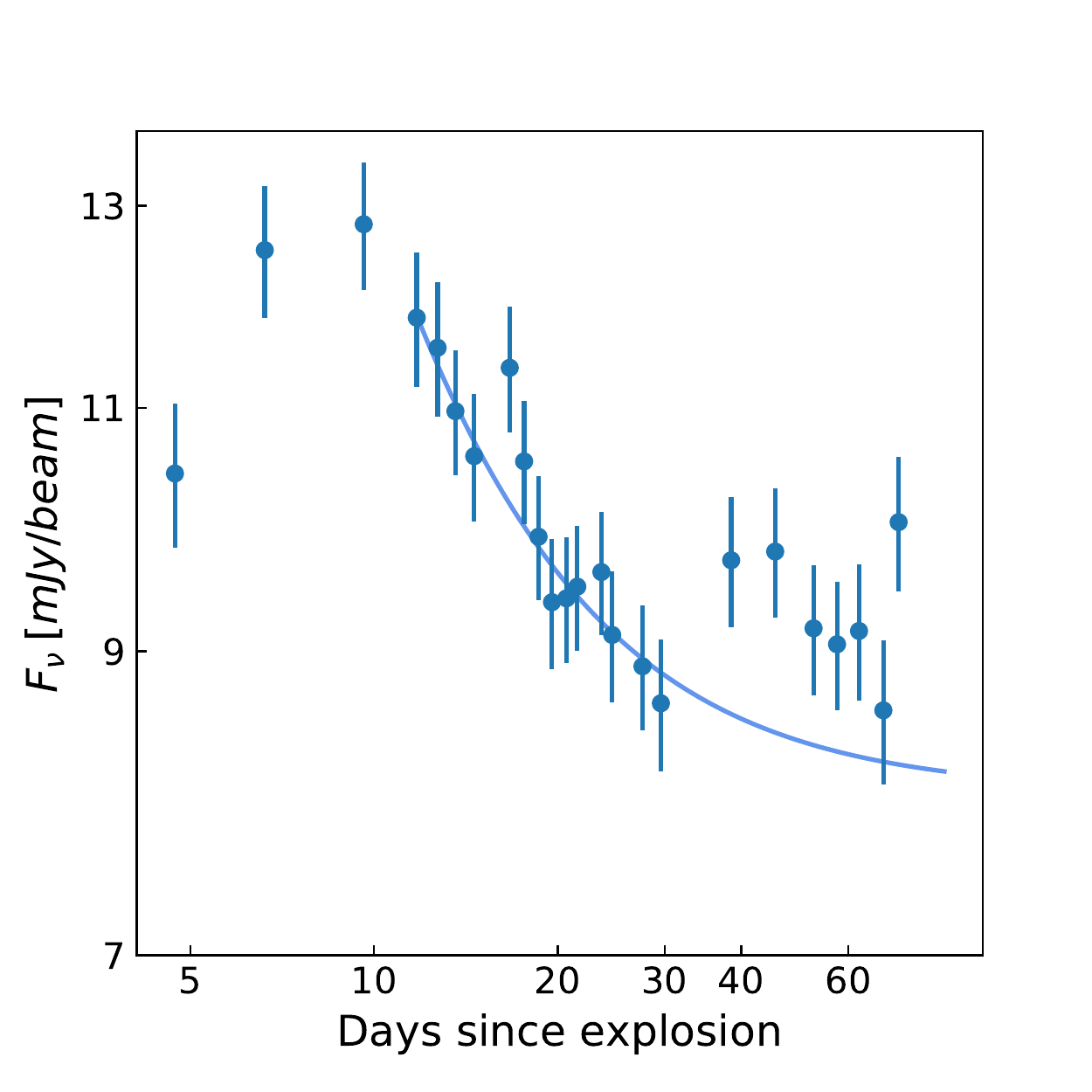}
\caption{\footnotesize{Radio emission as measured by AMI-LA and reported in Table \ref{Table: radio data}}. Also plotted is the results of the fitting process we conducted to estimate the constant underlying emission and described in $\S$\ref{sec:single_frequ}. The power law index of the flux with time, $b_{avg}=-1.66 \pm 0.28$, is an average of the power laws of the different K, Ka and Q sub-bands. The coefficient $A=233 \pm 54$ and the constant $C=8.0 \pm 1.1$\,mJy are product of the fit of the function $F_{\nu} = At^{b_{avg}} + C$ 
to the flux measured by AMI-LA in the optically thin regime.}
\label{fig:AMI light curve and fit}
\end{center}
\end{figure}

    In $\S$\ref{sec:background emission} we discussed the low angular resolution of AMI-LA and the resulting quiescent  underlying emission from the SN surrounding. To estimate this emission excess, we assume that the optically thin regime of AMI-LA light curve behaves as a power law in time plus a constant. To estimate the power law index we fit the data acquired on $11$ to $22$\,days after explosion, at K-, Ka- and Q-bands, to a power law function of time. We use only this time period since the flux measured by AMI-LA at later times is plateauing and probably represents the somewhat constant quiescent emission. In addition, as we show later in $\S$\ref{sec:spectral_index}, electron cooling is important in this time range while later it becomes negligible. We then average all these power laws to get an average power law index of $b_{avg}=-1.66 \pm 0.28$. We now fit the emission measured by AMI-LA in the optically thin regime on $11.76$ to $21.56$\,days after explosion, with the simple function $F_{\nu} = At^{b_{avg}} + C$. The result of this fit is $C=8.0 \pm 1.1$\,mJy ($\chi^2 _r= 0.23$, dof$=6$), and it is shown in Fig. \ref{fig:AMI light curve and fit}. Hence, we estimate the underlying constant emission evident in AMI-LA observations to be $8.0 \pm 1.1$\,mJy. Note, that the temporal power-law evolution is expected to change over time (as evident from fitting the temporal power-laws above in two different time periods) and the AMI fitting process does not take this fully into account, as we lack information about this varying evolution in the AMI band. Moreover, the flux from the quiescent sources that contaminate the AMI measurement, may experience slight variations that also effect the results here. 

    We subtract the constant emission based on the above estimate and fit Eq. 4 in \cite{chevalier_1998} to the $15.5$\,GHz light curve measured by AMI-LA, similar to the fit we performed to the e-MERLIN data. We use AMI-LA measurements from the first detection to $21.56$\,days after explosion only. The resulted power law index are $a=2.1 \pm 1.6$ and $b=-1.75 \pm 0.52$. The fitted peak flux at $15.5$\,GHz is $F_{\nu_a} = 5.33 \pm 1.61$\,mJy, at $8.39 \pm 1.49$\,days after explosion ($\chi^2 _r = 0.72$, dof$=7$). We used the peak at $15.5$\,GHz to calculate $R$, $B$, $v_{sh}$ and $\dot{M}$ (assuming $p=3$, $v_{sh} = R/t_a$ and $v_w=1000\,{\rm km\,s}^{-1}$) and find a rough (due to a large uncertainty) estimate of the shockwave velocity of $v_{sh}\approx 4.8 \pm 1.4\times 10^{4}\,{\rm km\,s}^{-1}$. While this velocity is somewhat higher than the velocities we derived earlier on, this result is consistent within $1\sigma$ of the previously measured velocities (see Table~\ref{Table: Peak fits}). Moreover, as we note above, the fitting of the AMI data should be treated with caution and so should any estimated property that is based on it.

\subsection{Broadband Spectrum Temporal Analysis}
\label{sec:multi_time}

The full time and spectral evolution of the self-absorbed synchrotron emission can be described by introducing a parameterized model as the one shown in Eq. 4 in \cite{chevalier_1998}. Below we perform a multi-frequency multi-epoch $\chi^2$ minimization fit of this model to the SN\,2020oi radio data. The free parameters in this process are the peak flux $F_{\nu_a} \left( t_a \right)$, its frequency $\nu_a$ and the spectral index $\beta$. The power laws of the light curve with time, $a$ (optically thick) and $b$ (optically thin), are also free parameters in this fitting process.

\begin{figure*}
\begin{center}
\hbox{\hspace{-6em}\includegraphics[width=1.25\linewidth]{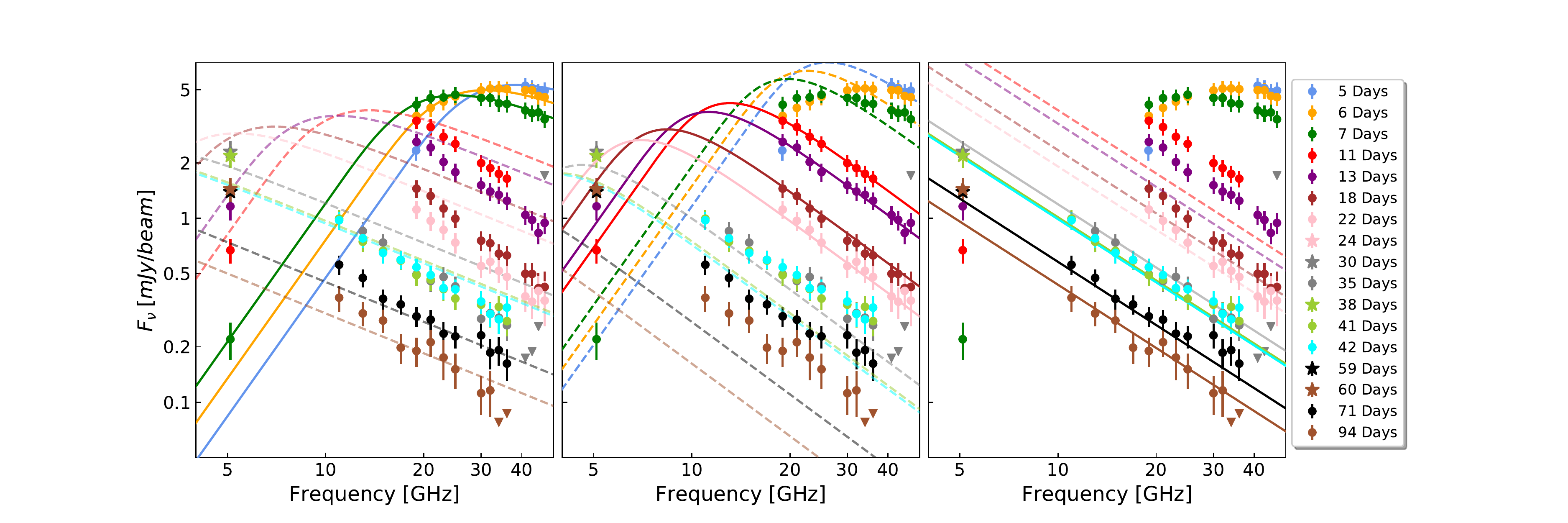}}
\caption{\footnotesize{The left and the middle panels are showing model fitting of Eq. $4$ in \cite{chevalier_1998} to the spectra of SN\,2020oi at different epochs. We fit the peak parameters $F_{\nu_a} \left( t_a \right)$ and $\nu_a$, the spectral slope $\beta$ and the time evolution slopes $a$ and $b$. In the left panel we assume $t_a=7$\,days while in the middle panel we assume $t_a=13$\,days. The left panel shows the results of fitting the first three epochs only (solid), and the extrapolated spectra (dashed) at later times. The middle panel shows the results of fitting the data taken $11$ to $24$\,days after explosion (solid), and the extrapolated spectra (dashed) at different times. The right panel shows the results of fitting a power law in frequency and time to the data starting $30$\,days after explosion. Since our late time observations do not constrain the radio peak we cannot make use of the model fitted above. The fitted spectra is shown in the right panel (solid), together with the extrapolated spectra (dashed) at early times.}\label{fig:VLA - Chevalier fits - with m}}
\end{center}
\end{figure*}

Since the full optically thick to optically thin spectrum seems to be captured only in the first three observing epochs, we first use only those epochs in a combined fit. Adopting $t_a=7$\,days, our best fit parameters are a peak flux of $4.38 \pm 0.62$\,mJy and frequency of $20.2 \pm 4.0$\,GHz. The spectral index is $\beta=-0.72 \pm 0.40$, while the power laws of the optically thick and thin regimes of the light curve are $a=3.0 \pm 1.0$ and $b=-1.4 \pm 0.9$, respectively. The resulted $\chi^2 _r$ in this case is $0.12$ (dof$=24$). The power laws of the temporal evolution have large uncertainties and therefor should be treated carefully. We then evolve this fitted model to future time and extrapolate the radio emission expected in our additional observing epochs. While our fit describes the first three epochs well, as shown in the left panel of Fig. \ref{fig:VLA - Chevalier fits - with m}, it does a poor job in describing the emission at later times. Also, the spectral index in later times is much steeper than the spectral index early on. However, as noted earlier (\S\,\ref{sec:single_epoch}), the shallow spectral index in the optically thin regime may be due to having data only in frequencies that are very close to the radio peak frequency. Thus, there is a good chance we are only witnessing the transition to the optically thin regime.

In $\S$\ref{sec:spectral_index} we argue that electron cooling is in effect in early times. However, it is most significant around the optical peak luminosity which is after the first three epochs. Additionally, we see some change in the spectral index behaviour between the VLA data obtained on days 11-22 and VLA data obtained afterwards. Thus, we next perform a fit to data that we obtained between $11$ to $24$\,days after explosion. Adopting $t_a=13$\,days here, our best fit parameters are a peak flux of $3.75 \pm 0.64$\,mJy and frequency of $10.6 \pm 1.4$\,GHz. The spectral index is $\beta=-1.31 \pm 0.44$, while the power laws of the optically thick and thin regimes of the light curve are $a=1.6 \pm 1.1$ and $b=-1.86 \pm 0.63$, respectively. The resulted $\chi^2 _r$ in this case is $0.26$ (dof$=41$). We note here, as in the previous fit, that the power laws of the temporal evolution have large uncertainties and therefor should be treated carefully. The middle panel of Fig. \ref{fig:VLA - Chevalier fits - with m} shows our fitted model, including extrapolations of the radio emission at epochs that are not used in the fitting process. As in the previous fit, the extrapolation of the current model to earlier and later times does not represent well the measurements at these times (see discussion below).

We next examine the late time emission by fitting flux measurements at times $\geq 35$\,days after explosion. 
Since our observations at these times are not constraining the peak we do not fit the same model we fitted above. Instead, we fit an optically thin emission which is described by a power law in frequency and time, i.e., $F_{\nu} \left( t \right) \sim \nu^{\beta} t^{b}$, where $\beta$ and $b$ are free parameters. Our modeling suggests $\beta = -1.14 \pm 0.14$ and $b = -1.02 \pm 0.09$ ($\chi^2 _r = 1.64$; dof$=62$). The right panel of Fig. \ref{fig:VLA - Chevalier fits - with m} shows our fitted model, including extrapolations of the radio emission model to epochs not used in the fitting process. We do not show the extrapolated lines of the first three epochs since they exhibit the transition from optically thick to optically thin emission. As the figure shows, while the data can be described quite well by the model at times $\geq35$\,days after explosion, the model prediction for earlier time emission deviates significantly from the observed radio emission. 

The rapid temporal evolution shown in the middle panel of Fig.~\ref{fig:VLA - Chevalier fits - with m} (between $11$ and $24$\,days) is expected as electron cooling is in effect. However, later on, the time evolution approaches $t^{-1}$ (as shown in the right panel of Fig.~\ref{fig:VLA - Chevalier fits - with m}) when the electron cooling no longer effects the emission in the observed frequencies. This is also the reason why the extrapolated radio emission from each fit over or under-predicts the observed emission in the time preceding or following the time period in which the fit was done. Since in addition to the temporal behaviour variations, there is a large uncertainty in the temporal power-law parameters in each fit, we refrain from using the combined temporal behaviour fits for estimating the shockwave parameters (e.g., shockwave radius) as the uncertainty of these estimates will be so large, it will render them useless.
Moreover, considering the single epoch modeling results together with the overall temporal behaviour of the decreasing peak flux points to a deviation from the standard simplistic model of a showckwave moving at a constant velocity in a spherical CSM structure with a density structure of $\rho_{csm}\propto r^{-2}$. This might be explained by a shockwave traveling in a more complex CSM density structure.  

\subsection{Electron Cooling}
\label{sec:spectral_index}

\begin{figure}
\begin{center}
\includegraphics[width=\linewidth]{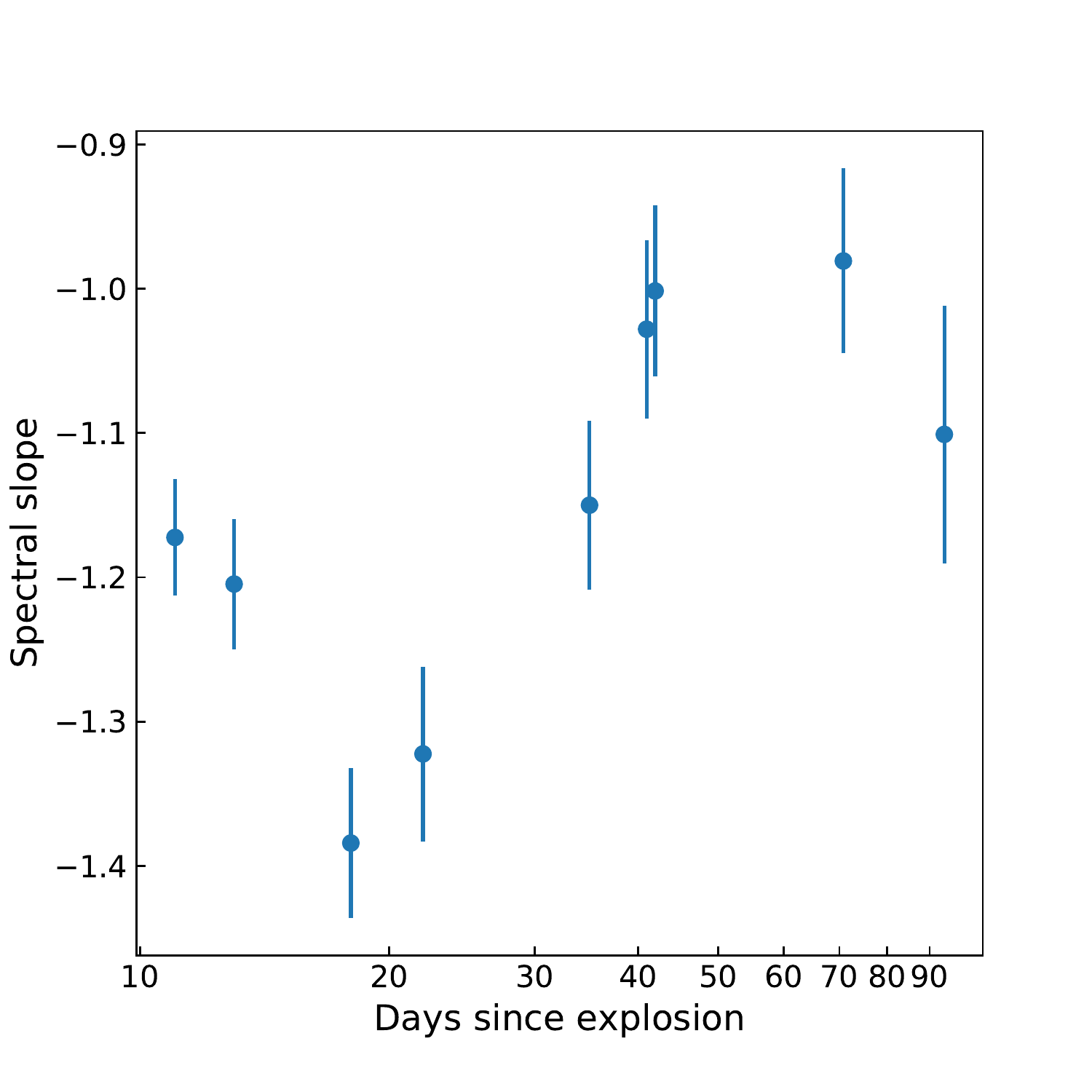}
\caption{\footnotesize{Spectral slope measured in different time by fitting a power law to the optically thin regime of the spectrum. Observations of Ku-band when the VLA was in D configuration, and of C- and lower frequency of X-bands when it was in C configuration, were not used to fit the power law, as described in $\S$\ref{sec:background emission}}}
\label{fig:VLA spectral slope}
\end{center}
\end{figure}

In the CSM interaction model described in $\S$\ref{sec:radio_analysis} we assumed a fixed electron energy power law of $p=3$, when translating the radio peak flux and frequency measurements to estimates of $R$ and $B$. We already saw in the previous section that the spectral index deviates from the expected value of $\beta=-1$ (if $p=3$ and $\beta=-[p-1]/2$). We further test this by measuring the spectral slope $\beta$ at the optically thin regime at different times, starting from day $11$ after explosion. 
As discussed in $\S$\ref{sec:background emission}, the observations in Ku-band when the VLA was in D configuration, and in C-band when it was in C configuration, are suffering from flux contamination. Therefore, we removed these observations when fitting the power law to the optically thin regime. The evolution of the spectral slope in time is shown in Fig. \ref{fig:VLA spectral slope}.

As shown in the figure, the spectral index varies with time. Between $11$ to $22$\,days the spectral index is steeper than the expected value of $\beta=-1$ (a similar behaviour was observed in SN\,2012aw; \citealt{yadav_2014}). A possible explanation for this behaviour is electron cooling, either due to Synchrotron cooling or inverse Compton cooling (see a discussion in \citealt{Bjornsson_2004}). In these two former scenarios, if the cooling timescale is shorter then the adiabatic timescale, then the flux above a certain cooling frequency is reduced compared to the non cooling SSA only model, effectively leading to a steeper spectral index. Also, it seems that after $\sim 40$\,days, the spectral index settles onto a value of $\beta \approx -1$.

The synchrotron cooling frequency is
\begin{equation}
\nu_{syn\_cool} =  \frac{18\pi m_{e} c q_{e}}{\sigma_{T}^2 B^{3} t^{2}}, 
\end{equation}
where $m_e$ is the electron mass, $q_e$ is a unit charge, and $\sigma_{T}$ is the Thomson cross section. Using this equation with the values of the magnetic field we found in \S~\ref{sec:single_epoch}, we find that the synchrotron cooling frequency is $\nu_{syn\_cool}>200$\,GHz at all times and thus does not effect the radio emission at the observed frequencies. It is more probable then that inverse Compton cooling is the dominant process here that leads to the steep spectral index we observe. 

In the inverse Compton case, the cooling frequency is evolving as follows: 
\begin{align}
    \nonumber \nu_{comp} \approx  & ~0.324 \left( \frac{L_{bol}}{2\times 10^{42}\,{\rm erg \, s}^{-1}} \right) ^{-2} \left( \frac{\epsilon_{B}}{0.1} \right)^{1/2}\\ &\left(\frac{\dot{ M}\,[10^{-6}\,{M}_{\odot}\,{\rm yr}^{-1}]}{v_{w} [10\,{\rm km}\,{\rm s}^{-1}]}\right)^{1/2}  \left(\frac{v_{sh}}{10^{4}\,{\rm km}\,{\rm s}^{-1}}\right)^{4} \, {\rm GHz}
\end{align}
Note that the cooling frequency depends on $\epsilon_{B}$ and in addition the estimates of $v_{sh}$ and $\dot{M}$ using the radio peak data also depends on the ratio between the micro-physical parameters $(\epsilon_{e}/\epsilon_{B})$. A constraint on the inverse Compton cooling frequency can thus be translated into a constraint on $\epsilon_{B}$, assuming a value for $\epsilon_{e}$. In the case of SN\,2020oi, we assume that the Compton cooling frequency travels above our observed range $\geq 40$\,GHz at roughly $30-40$\,days after explosion. Using this in combination with the observed bolometric luminosity estimate at that time $L_{bol}\sim 2.7\times 10^{41}$\,erg\,s$^{-1}$, and also assuming $\epsilon_{e}\approx 0.1$, results in the limit $\epsilon_{B} \lesssim 0.0005$. Thus we find that there is deviation from equipartition with $(\epsilon_{e}/\epsilon_{B})\gtrsim 200$. This deviation from equipartition is similar to the one  found in both SN\,2012aw and SN\,2013df. 

We next estimate the expected X-ray emission as a result of the inverse Compton process. According to Eq. $32$ in \cite{chevalier_fransson_2006}, and adopting a bolometric luminosity of $\sim 2.3\times 10^{42}$\,erg\,s$^{-1}$, the expected inverse Compton X-ray emission is $\approx  1.2\times 10^{39} \,{\rm erg\,s}^{-1}$. As one can see, this is below our observed limit of $L_{x}< 10^{40} \,{\rm erg\,s}^{-1}$ (\S\,\ref{sec:xray_obs}). Thus, unfortunately, deriving any additional constraints, using the X-ray observations, is not possible here. 

\subsection{The effect on non-equipartition on shockwave parameter estimates}
\label{sec:non_eq}

The shockwave parameter estimates of $R_{p}$ and $B_{p}$ and the derived estimates of $v_{sh}$ and $\dot{M}/v_{w}$ in \S~\ref{sec:single_epoch} are based on the equipartition assumption. The derived values of these parameters will change once including the deviation from equipartition that we find. Adopting $(\epsilon_{e}/\epsilon_{b})\gtrsim 200$ will result in the reduction of the shockwave radius values by $\gtrsim 24\%$ (and the shockwave velocity accordingly), the reduction the magnetic field strength value by $\gtrsim 67\%$, and the increase in the $\dot{M}/v_{w}$ value by a factor of $\gtrsim 12$.

\section{Conclusions and Summary}
\label{sec:conclusions}

Here we report the early optical discovery of SN\,2020oi and present a detailed panchromatic measurement set of the SN. In the optical we find that SN\,2020oi is a normal young Type Ic SN with early photospheric velocity of $\sim 15,000 \,{\rm km\,s}^{-1}$. The series of optical spectra that we present shows a typical evolution of a stripped envelope SN. In the X-ray, observations undertaken by the {\it Swift} satellite did not reveal any bright X-ray SN emission. However, the X-ray observations sensitivity is limited by the bright existing background emission from the host galaxy. Thus the X-ray observations only provide a weak constraint of $L_{X}\leq 10^{40}\,{\rm erg\,s}^{-1}$. In the radio a bright mJy source is detected in observation undertaken by several facilities. 

The radio observations we present in this paper were undertaken by the VLA, ATCA, e-MERLIN and the AMI-LA telescopes. The observations resulted in multi-epoch multi-frequency detailed measurements. We analyse these measurements in several ways assuming a single shockwave model driven by the interaction of the SN ejecta with the CSM \citep{chevalier_1998}. We performed modeling of the radio data in several ways including: single epoch spectral modeling, single frequency modeling and spectral multi epoch modeling. 

Our modeling of the radio data points towards a non-equipartition shockwave traveling in a dense CSM environment. We find that on average the shockwave is moving at a constant velocity, although a standard constant velocity shockwave model alone fail to reproduce the full data set. This may be explained by a slight deviation of the CSM density structure from a $r^{-2}$ power-law function. If this is indeed the case, this may suggest that the mass-loss rate had been slowly changing. However, the lack of detailed high-resolution data at low GHz frequencies  limits the analysis performed here, and does not allow a more complex modeling.

Our radio dataset also exhibit a period in which the spectral index in the optically thin regime become rather steep. A possible explanation is the effect of electron cooling by the inverse Compton process on the observed spectrum. After about $40$\,days, the spectral index becomes shallower and reached a value of $\beta \approx -1$, which is the typical spectral index observed in stripped envelope SN when cooling is not in play. We use the departure of the inverse Compton cooling from our observing bands at $\sim 40$\,days to estimate the ratio between two key microphysical parameters, $\epsilon_e/\epsilon_B \gtrsim 200$. Also, the relation between the spectral index $\beta$ and the electron energy distribution power-law index $p$, $\beta=-(p-1)/2$ is valid in the absence of electron cooling, thus after $\sim 40$\,days. This points to an index of $p\approx 3$, which is the typical index observed in Type Ic SNe \citep{chevalier_fransson_2006}.

Large deviations from equipartition in SN-CSM shockwaves have been observed in the past in several cases (e.g SN\,2011dh \citealt{soderberg_2012, horesh_mnras_2013}; SN\,2012aw \citealt{yadav_2014}; SN\,2013df \citealt{kamble_2016}), although in other cases the shockwave was found to be in equipartition (e.g., \citealt{Bjornsson_2004}). Early high cadence panchromatic observations played a key role in identifying these deviations. In many other cases, there is not enough information to determine whether there is a deviation from equipartition. In these cases, the derived shockwave and CSM parameters may not truly represent their real values. Here, for example, the shockwave velocity estimate is lowered from $\approx 4\times 10^{4} \, {\rm km\,s}^{-1}$ to $\approx 3\times 10^{4} \, {\rm km\,s}^{-1}$ when taking into account the deviation from equipartition. As for the mass-loss rate estimate, the effect on it is much greater, and in our case it increases by a factor of $>12!$ The question why some SN shockwaves exhibit equipartition while other show large deviations from equiparition still remains an open question. Before attempting to answer this question, a better characterization of a large sample of SNe at early times. 

Overall, SN\,2020oi is a normal Type Ic SN in optical wavebands, with a somewhat non standard evolution of its radio emission. The SN-CSM shockwave we find in our analysis suggest velocities in the range of $3-4 \times 10^{4} \, {\rm km\,s}^{-1}$, which is typical of Type Ic SNe. The mass-loss rate we deduce including the deviation from equipartition is on the higher end of the mass-loss rate in stripped envelope SN, but not in any extreme way (\citealt{smith_2014}). Detailed panchromatic observational campaigns, such as the one undertaken here, are required to build a large sample of well-characterized stripped envelope SNe that may be used to search for answers to some of the open questions in the field of SNe.

\section*{Acknowledgments}
\label{sec:ACK}
A.H. is grateful for the support by grants from the Israel Science Foundation, the US-Israel Binational Science Foundation (BSF), and the I-CORE Program of the Planning and Budgeting Committee and the Israel Science Foundation. 
T.M. acknowledges the support of the Australian Research Council through grant FT150100099.
D.D. is supported by an Australian Government Research Training Program Scholarship. 
D.R.A.W was supported by the Oxford Centre for Astrophysical Surveys, which is funded through generous support from the Hintze Family Charitable Foundation.
A.~A.~Miller is funded by the Large Synoptic Survey Telescope Corporation, the
Brinson Foundation, and the Moore Foundation in support of the LSSTC Data
Science Fellowship Program; he also receives support as a CIERA Fellow by the
CIERA Postdoctoral Fellowship Program (Center for Interdisciplinary
Exploration and Research in Astrophysics, Northwestern University).
M.P.T acknowledges financial support from the State
Agency for Research of the Spanish MCIU through the "Center of
Excellence Severo Ochoa" award to the Instituto de Astrofísica de
Andalucía (SEV-2017-0709) and through grant PGC2018-098915-B-C21
(MCI/AEI/FEDER, UE).
J.M. acknowledges financial support from the State Agency for Research of the Spanish MCIU through the ``Center of Excellence Severo Ochoa'' award to the Instituto de Astrof\'isica de Andaluc\'ia (SEV-2017-0709) and from the grant RTI2018-096228-B-C31 (MICIU/FEDER, EU).
A.G.Y’s research is supported by the EU via ERC grant No. 725161, the ISF GW excellence center, an IMOS space infrastructure grant and BSF/Transformative and GIF grants, as well as The Benoziyo Endowment Fund for the Advancement of Science, the Deloro Institute for Advanced Research in Space and Optics, The Veronika A. Rabl Physics Discretionary Fund, Paul and Tina Gardner, Yeda-Sela and the WIS-CIT joint research grant;  AGY is the recipient of the Helen and Martin Kimmel Award for Innovative Investigation.
M.R. has received funding from the European Research Council (ERC) under the European Union's Horizon 2020 research and innovation programme (grant agreement n759194 - USNAC). 
M.~W.~Coughlin acknowledges support from the National Science Foundation with grant number PHY-2010970. C.F.~gratefully acknowledges support of his research by the Heising-Simons Foundation (\#2018-0907).
Based on observations obtained with the Samuel Oschin Telescope 48-inch and the 60-inch Telescope at the Palomar Observatory as part of the Zwicky Transient Facility project. ZTF is supported by the National Science Foundation under Grant No. AST-1440341 and a collaboration including Caltech, IPAC, the Weizmann Institute for Science, the Oskar Klein Center at Stockholm University, the University of Maryland, the University of Washington, Deutsches Elektronen- Synchrotron and Humboldt University, Los Alamos National Laboratories, the TANGO Consortium of Taiwan, the University of Wisconsin at Milwaukee, and Lawrence Berkeley National Laboratories. Operations are conducted by COO, IPAC, and UW. Partly based on observations made with the Nordic Optical Telescope. SED Machine is based upon work supported by the National Science Foundation under Grant No. 1106171. The National Radio Astronomy Observatory is a facility of the National Science Foundation operated under cooperative agreement by Associated Universities, Inc. The Australia Telescope Compact Array is part of the Australia Telescope National Facility which is funded by the Australian Government for operation as a National Facility managed by CSIRO. We acknowledge the Gomeroi people as the traditional owners of the Observatory site. e-MERLIN is a National Facility operated by the University of Manchester at Jodrell Bank Observatory on behalf of STFC. We thank the staff of the Mullard Radio Astronomy Observatory
for their assistance in the commissioning, maintenance and operation of
AMI, which is supported by the Universities of Cambridge and Oxford.
We also acknowledge support from the European Research Council under
grant ERC-2012-StG-307215 LODESTONE. 

\bibliography{SN2020oi.bib}

\newpage

\end{document}